\documentclass[%
 reprint,
 amsmath,amssymb,
 aps,
]{revtex4-1}

\usepackage{graphicx}
\usepackage{dcolumn}
\usepackage{bm}

\begin{document}


\title{Large-scale shell-model calculations for unnatural-parity high-spin states 
in neutron-rich Cr and Fe isotopes}

\author{Tomoaki Togashi$^{1}$}
 \email{togashi@cns.s.u-tokyo.ac.jp}
\author{Noritaka Shimizu$^{1}$}
\author{Yutaka Utsuno$^{1,2}$}
\author{Takaharu Otsuka$^{1,3,4}$}
\author{Michio Honma$^{5}$}
\affiliation{
$^{1}$Center for Nuclear Study, University of Tokyo, Hongo, Bunkyo-ku, Tokyo, 113-0033, Japan\\
$^{2}$Advanced Science Research Center, Japan Atomic Energy Agency, Tokai, Ibaraki, 319-1195, Japan\\
$^{3}$Department of Physics, University of Tokyo, Hongo, Bunkyo-ku, Tokyo, 113-0033, Japan\\
$^{4}$National Superconducting Cyclotron Laboratory, Michigan State University, East Lansing, Michigan, 48824, USA\\
$^{5}$Center for Mathematical Sciences, University of Aizu, ikki-machi, Aizu-Wakamatsu, Fukushima, 965-8580, Japan
}

\date{\today}

\begin{abstract}
We investigate unnatural-parity high-spin states in neutron-rich Cr 
and Fe isotopes using large-scale shell-model calculations. 
These shell-model calculations are carried out within the model space of 
$fp$-shell + $0g_{9/2}$ + $1d_{5/2}$ orbits with the truncation 
allowing $1\hbar\omega$ excitation of a neutron. 
The effective Hamiltonian consists of GXPF1Br for $fp$-shell orbits 
and $V_{\rm MU}$ with a modification for the other parts. 
The present shell-model calculations can describe and predict 
the energy levels of both natural- and unnatural-parity states up to 
the high-spin states in Cr and Fe isotopes with $N\le35$. 
The total energy surfaces present the prolate deformations on the whole 
and indicate that the excitation of one neutron into the $0g_{9/2}$ orbit 
plays the role of enhancing the prolate deformation. 
For the positive(unnatural)-parity states in odd-mass Cr and Fe 
isotopes, their energy levels and prolate deformations 
indicate the decoupling limit of the particle-plus-rotor model. 
The sharp drop of the $9/2_{1}^{+}$ levels in going from $N=29$ to $N=35$ 
in odd-mass Cr and Fe isotopes is explained by the Fermi surface approaching 
the $\nu 0g_{9/2}$ orbit.
\end{abstract}

\maketitle


\section{Introduction}

The evolution of the shell structure in neutron-rich nuclei is one of the 
main interests of modern nuclear physics. 
The neutron-rich $fp$-shell region has attracted particular attraction 
in this context because new neutron magic numbers have recently 
been established. 
The subshell closure at $N=32$ is indicated 
from high $2_{1}^{+}$ levels and 
reduced $B(E2,0_{1}^{+} \rightarrow 2_{1}^{+})$ values in 
$^{52}$Ca \cite{ca52_1,ca52_2}, $^{54}$Ti \cite{ti54_1,ti54_2}, and 
$^{56}$Cr \cite{cr56_1,cr56_2} compared to neighboring isotopes. 
A new subshell closure at $N=34$ was 
observed in $^{54}$Ca in 2013 \cite{gxpf1br}, 
more than a decade after its prediction \cite{magic}. 
Evaluating the evolution of those shell gaps is facilitated by 
advances in shell-model calculations for the full $fp$-shell
space \cite{gxpf1, Caurier_RMP}.  
The appearance of the $N=32$ and $N=34$ magic numbers 
in the proton-deficient region is attributed 
to the tensor-force driven shell evolution
\cite{tensor}. 

The evolution of the shell gap at $N=40$, one of the main objectives 
of this paper, constitutes a key to understanding 
the characteristic features of the nuclear structure in the 
neutron-rich $N\sim 40$ region. 
While the $N=40$ magicity seems to be rather stable in Ni isotopes 
\cite{ni68_broda, ni68_ishii, ni68_sorlin}, 
it breaks down completely in Cr and Fe 
isotopes: very low $2_{1}^{+}$ levels and 
large $B(E2,0_{1}^{+} \rightarrow 2_{1}^{+})$ 
values in $^{60-64}$Cr \cite{cr_fe_2p,cr60-62_1,cr60-62_2,cr64_fe68} 
and $^{66-68}$Fe \cite{cr_fe_2p,fe66,cr64_fe68} 
cannot be reproduced without considerable neutron excitations  
across the $N=40$ shell gap \cite{Kaneko, Oba, Lenzi}. 
Most recently, it has been reported that the breakdown of the $N=40$ magicity 
can possibly be extended even to Ti isotopes \cite{ti} 
having less proton-neutron quadrupole collectivity than the Cr and Fe isotopes. 
This abrupt nuclear-structure change from Ni to lower-$Z$ isotopes 
is analogous to what is observed in the so-called ``island of inversion'' 
region around $^{32}$Mg \cite{island} 
where neutron excitation induces a large deformation. 
The evolution of the $N=40$ shell gap 
contributes greatly to the formation of the island of inversion in Cr and Fe 
isotopes. Indeed, it is suggested in \cite{Lenzi} that the $N=40$ shell 
gap reduces with a decreasing proton number in a similar way to 
the evolution of the $N=20$ gap proposed in \cite{n20mcsm}. 
The tensor force accounts for this reduction of the harmonic-oscillator 
shell gaps at $N=20$ and 40 toward lower $Z$, 
and also plays a crucial role in the occurrence of 
the spherical-oblate-prolate shape coexistence in $^{68}$Ni 
through the Type-II shell evolution proposed in \cite{Tsunoda}. 

While the evolution of the $N=40$ shell gap thus causes a number of 
intriguing phenomena, it is very difficult to directly deduce this shell gap 
from the structure of the Cr and Fe isotopes that belong to the island of
inversion. 
This is because low-lying states of those nuclei are dominated 
by multi-particle multi-hole excitations, and this property is not very 
sensitive to small changes of the shell gap. 
On the other hand, the states dominated by 
one-particle one-hole ($1p$-$1h$) excitation 
provide more direct information on the $N=40$ shell gap, 
when those levels are compared to the $0p$-$0h$ states. 
For Cr and Fe isotopes, low-lying unnatural-parity states 
with $N\le 35$ are regarded as such $1p$-$1h$ states 
because low-lying natural-parity states 
in this region are excellently described within 
the $fp$-shell configurations \cite{gxpf1}. 
The nuclear properties of unnatural-parity states in 
Cr and Fe isotopes in the last decade have been measured mainly 
from $\gamma$-ray spectroscopy experiments. 
Important findings from recent experiments 
include a sharp drop of the $9/2_{1}^{+}$ levels in going from 
$N=29$ to $N=35$ observed in Cr and Fe isotopes 
\cite{cr55,cr57,cr59,fe59-60,fe61}.
If this change in energy level is an indication 
of a sharp evolution of the $0g_{9/2}$ orbit with an increasing neutron
number as suggested in \cite{Kaneko}, 
the driving force of the evolution needs to be clarified.

Another significant nuclear-structure issue in the unnatural-parity states 
in Cr and Fe isotopes is their nuclear shapes; 
in particular, how these shapes are polarized by an additional neutron 
in the $0g_{9/2}$ orbit.
The development of deformation is indicated experimentally 
from the regular spacing between $\Delta I=2$ energy levels 
observed up to high-spin states \cite{cr55,cr55_2,cr57,fe59-60,fe61}. 
While various models show the dominance of prolate deformation in 
the ground states of Cr isotopes (see \cite{Lalazissis} for instance), 
an oblate deformation in the unnatural-parity band of $^{59}$Cr 
has been suggested in \cite{cr59}
from the observed isomeric $9/2^+$ state located at 503 keV, 
which should be the lowest among unnatural-parity states. 
This level is interpreted naively as 
the bandhead of the $K=9/2$ intrinsic state, 
which appears to be the lowest for oblate deformation. 
On the other hand, a recent study with a projected shell-model calculation 
has shown that prolate deformation is favored based on comparison 
with experimental data \cite{cr59_def}. 
A similar isomeric $9/2^+$ state is observed in $^{61}$Fe, and 
its magnetic moment \cite{fe61_g} and absolute value of quadrupole 
moment \cite{fe61_sqm} support prolate deformation 
according to the particle-triaxial-rotor model \cite{fe61}. 
More elaborate systematic calculations incorporating various degrees of freedom 
for deformation and configuration help draw a definite conclusion about 
the shape evolution in Cr and Fe isotopes. 

In this paper, we report on systematic large-scale shell-model calculations 
for natural-parity and unnatural-parity states in Cr and Fe isotopes for
$N\le 35$, by clarifying the structure of their $0p$-$0h$ and 
$1p$-$1h$ bands. 
Using a combination of known effective interactions that are slightly
modified, we achieve excellent agreement with experiment 
in the whole region studied. 
This highly descriptive power enables extracting shell and shape 
evolutions in this region. 
This paper is organized as follows. In Sec.~\ref{sec:framework}, we present 
our theoretical framework including the effective Hamiltonian to be used 
in this work. 
In Sec.~\ref{sec:results}, first, we show the results of the energy levels 
for neutron-rich Cr and Fe isotopes; secondly, we discuss the deformation 
for the natural- and unnatural-parity states in these nuclei.
In Sec.~\ref{sec:n0g9}, we discuss how the $\nu0g_{9/2}$ orbit evolves 
in this mass region by using theoretical and experimental information. 
Conclusions of this study are given in Sec.~\ref{sec:conclusion}.

\section{\label{sec:framework}Theoretical framework}

\subsection{Shell-model calculation}

In the mass region of $Z$ or $N$ from $20$ to $40$, the natural-parity states 
with $\pi=+(-)$ for even(odd)-mass nuclei are usually described well within 
the $fp$-shell model space with the $^{40}$Ca inert core. 
However, the unnatural-parity states with $\pi=-(+)$ for even(odd)-mass nuclei 
cannot be described within one major $fp$ shell. 
In the quasi-SU3 model \cite{q_SU3}, the pairs of $\Delta j=2$ orbits 
of a major shell describe quadrupole collectivity. 
Hence, we choose the $0g_{9/2}$ and $1d_{5/2}$ orbits and 
add them to the $fp$-shell model space. 
In the present shell-model calculation, the model space is composed of 
$fp$-shell ($0f_{7/2}$, $0f_{5/2}$, $1p_{3/2}$, $1p_{1/2}$) 
+ $0g_{9/2}$ + $1d_{5/2}$ orbits. 
To focus on the states dominated by $1p$-$1h$ excitation across 
the $N=40$ shell gap, a truncation is introduced so that one neutron 
is allowed to occupy the $0g_{9/2}$ or $1d_{5/2}$ orbit, 
which means that the $1\hbar\omega$ excitation of a neutron to the $gds$
shell occurs in only unnatural-parity states.

The present shell-model Hamiltonian has the following form which consists of 
one- and two-body terms:
\begin{eqnarray}
H = \sum_{i} \varepsilon_{i} c_{i}^{\dagger}c_{i}
+ \sum_{i<j,k<l} V_{ijkl}\,c_{i}^{\dagger}c_{j}^{\dagger}c_{l}c_{k}
+ \beta_{c.m.} H_{c.m.},
\label{eq:hmat}
\end{eqnarray}
where $\varepsilon_{i}$ and $V_{ijkl}$ represent the bare single-particle 
energy and the two-body matrix element, respectively, 
$c^{\dagger}_{i}(c_{i})$ denotes a creation(annihilation) operator of 
a nucleon in a single-particle orbit $i$, 
and $\beta_{c.m.} H_{c.m.}$ is the term proposed by Gloeckner and Lawson 
to remove the spurious center-of-mass motion due to the excitation 
beyond one major shell \cite{Lawson}. 
The parameter of $\beta_{c.m.}$ is applied as 
$\beta_{c.m.}\hbar\omega/A=10$ MeV, where $A$ is the mass number: $A=Z+N$. 
We take $\hbar\omega$ from the empirical value: $\hbar\omega=41.0A^{-1/3}$ MeV. 

In the present work, the Hamiltonian matrix with an $M$ scheme is diagonalized 
by the thick-restart Lanczos method with MSHELL64 code \cite{MSHELL64} 
for the dimension of the matrix below $10^{9}$ and KSHELL code \cite{KSHELL} 
for the dimension over $10^{9}$.

\subsection{\label{sec:effh}The effective Hamiltonian}

In this subsection, we explain the details of the effective Hamiltonian in 
Eq.~(\ref{eq:hmat}) in the present work. 
For the shell-model calculations within the $fp$-shell model space, 
the GXPF1A Hamiltonian \cite{gxpf1a} is often applied to reproduce and 
to predict the properties of nuclei in the region of $Z$ or $N$ 
from $20$ to $40$. Recently, the GXPF1A Hamiltonian has been improved 
to explain the observed energy levels in neutron-rich Ca isotopes: 
GXPF1B \cite{gxpf1b} with the modification of five $T=1$ two-body matrix 
elements and the bare single-particle energy which involve the $1p_{1/2}$ 
orbit from GXPF1A, and GXPF1Br \cite{gxpf1br} with the modification of 
the monopole interaction for $\langle 0f_{5/2}1p_{3/2}|V|0f_{5/2}1p_{3/2} 
\rangle_{T=1}$ from GXPF1B. The GXPF1B Hamiltonian can reproduce 
the energy levels of $^{51,52}$Ca, and the GXPF1Br Hamiltonian can reproduce 
those of $^{53,54}$Ca besides $^{51,52}$Ca. 
Here, we adopt the newest effective Hamiltonian, GXPF1Br, 
for $fp$-shell orbits. 
For neutron $1\hbar\omega$ excitation to the $0g_{9/2}$ or $1d_{5/2}$ 
orbits, we create the cross-shell two-body interaction between 
the $fp$-shell and $gds$-shell orbits from $V_{\rm MU}$ \cite{vmu}, 
which has the central force of a simple Gaussian form and the tensor force of 
the $\pi+\rho$ mesons exchange force to reproduce the monopole properties of 
the shell-model Hamiltonian in $sd$ and $fp$ shells. 
We utilize $V_{\rm MU}$ with the refinement \cite{vmu_r} 
by adding the M3Y spin-orbit force \cite{M3Y} and the density dependence 
for the central force described in \cite{Brown}. 
All two-body matrix elements are scaled by 
$(A/42)^{-0.3}$ as the mass dependence.

In addition to the cross-shell two-body interaction, 
the bare single-particle energies with the $^{40}$Ca inert core 
for the $0g_{9/2}$ and $1d_{5/2}$ orbits need to be determined. 
The bare single-particle energy of $0g_{9/2}$ is 
determined so as to reproduce the excitation energy of $9/2_{1}^{+}$ states 
systematically. However, determining the bare single-particle energy of 
$1d_{5/2}$ is more difficult than that of $0g_{9/2}$. 
The single-particle property of the $1d_{5/2}$ orbit in this 
mass region is masked. In fact, the experimental spectroscopic factors 
of the $5/2_{1}^{+}$ state in neutron-rich odd-mass Cr and Fe nuclei are 
rather small in comparison with those of the $9/2_{1}^{+}$ state 
\cite{NDS_A55,NDS_A57,NDS_A59}. 
Here, we assume that the bare single-particle energy of 
$1d_{5/2}$ is equal to that of $0g_{9/2}$. 
This assumption is considered to be reasonable 
because the splitting of the effective single-particle 
energies between $0g_{9/2}$ and $1d_{5/2}$ for nuclei taken in this study, 
around 2 MeV, is close to the phenomenological $gds$-shell orbits splitting 
in this mass region \cite{Bohr}. 
In the present work, the bare single-particle energy of $0g_{9/2}$ 
(= that of $1d_{5/2}$) is determined so as to 
be fitted to reproduce the excitation energies of $9/2_{1}^{+}$ in
$^{55,57,59}$Cr. The resulting value is $0.793$ MeV.

\begin{figure}[h]
\begin{tabular}{c}
  \includegraphics[width=8.0cm]{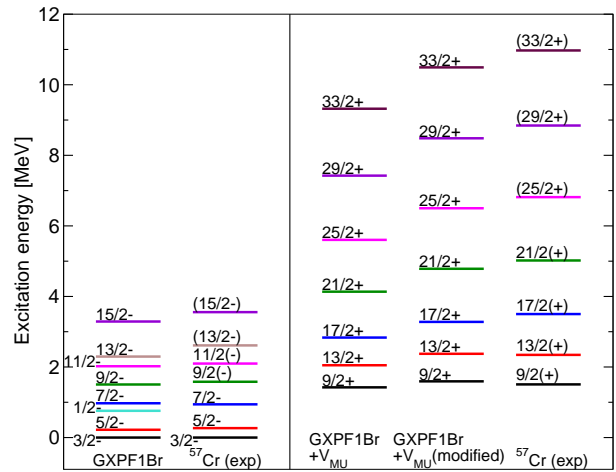}\\
\end{tabular}
\caption{\label{fig:eff_v}(Color online) Shell-model results and 
experimental data of the energy levels of $^{57}$Cr. 
The experimental data labeled ``$^{57}$Cr(exp)'' are taken from \cite{cr57}.}
\end{figure}

Figure~\ref{fig:eff_v} shows the energy levels of $^{57}$Cr and the results 
of the shell-model calculations with the above mentioned Hamiltonian. 
As seen in Fig.~\ref{fig:eff_v}, the $fp$ shell-model calculation 
with GXPF1Br reproduces the negative(natural)-parity states quite well 
and GXPF1Br + $V_{\rm MU}$ without modification reproduces 
the positive(unnatural)-parity states reasonably well. 
For further improvement, we change the cross-shell two-body matrix elements 
by $-1.0$ MeV for 
$\langle 0g_{9/2}0f_{5/2}\,|V|\,0g_{9/2}0f_{5/2} \rangle_{J=2,3,\,T=1}$ 
and by $+1.0$ MeV for 
$\langle 0g_{9/2}0f_{5/2}\,|V|\,0g_{9/2}0f_{5/2} \rangle_{J=6,7,\,T=1}$. 
This modification increases the energy intervals in the positive-parity 
band by about $15\%$ as a whole. 
These results labeled GXPF1Br + $V_{\rm MU}$(modified) 
in Fig.~\ref{fig:eff_v} agree with 
the experimental data within $\sim$500 keV. 
In the following, we present the results for 
Cr and Fe isotopes with the new effective Hamiltonian, GXPF1Br + 
$V_{\rm MU}$(modified). 
\begin{figure*}
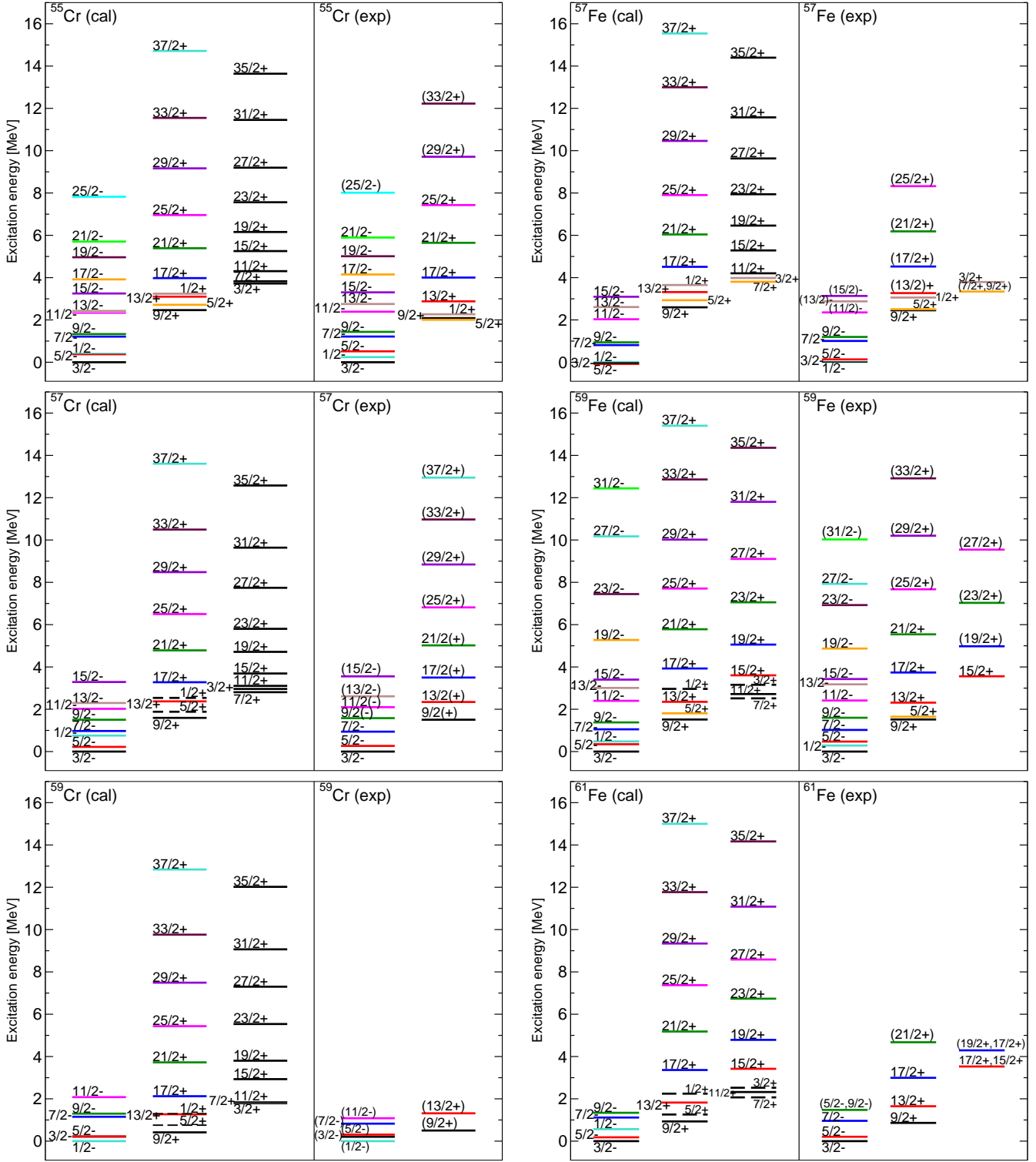

\begin{tabular}{c}
\begin{minipage}{0.5\hsize}
  \begin{center}
  \includegraphics[width=8.6cm]{55Cr_lv.eps}
  \end{center}
\end{minipage}
\begin{minipage}{0.5\hsize}
  \begin{center}
  \includegraphics[width=8.6cm]{57Fe_lv.eps}
  \end{center}
\end{minipage}
\vspace{1mm}
\end{tabular}
\begin{tabular}{c}
\begin{minipage}{0.5\hsize}
  \begin{center}
  \includegraphics[width=8.6cm]{57Cr_lv.eps}
  \end{center}
\end{minipage}
\begin{minipage}{0.5\hsize}
  \begin{center}
  \includegraphics[width=8.6cm]{59Fe_lv.eps}
  \end{center}
\end{minipage}
\vspace{1mm}
\end{tabular}
\begin{tabular}{c}
\begin{minipage}{0.5\hsize}
  \begin{center}
  \includegraphics[width=8.6cm]{59Cr_lv.eps}
  \end{center}
\end{minipage}
\begin{minipage}{0.5\hsize}
  \begin{center}
  \includegraphics[width=8.6cm]{61Fe_lv.eps}
  \end{center}
\end{minipage}
\end{tabular}
\caption{\label{fig:elv_odd}(Color online) Calculated and experimental energy 
levels of the lowest states for each spin and parity 
in odd-mass Cr and Fe isotopes. 
In each panel, the left and right present the calculated and 
experimental energy levels, respectively. 
The experimental data are taken from 
\cite{cr55_2,cr57,cr59,fe59-60,fe61,cr56_cr58,NDS_A55,NDS_A57,NDS_A59}.}
\end{figure*}
\begin{figure*}
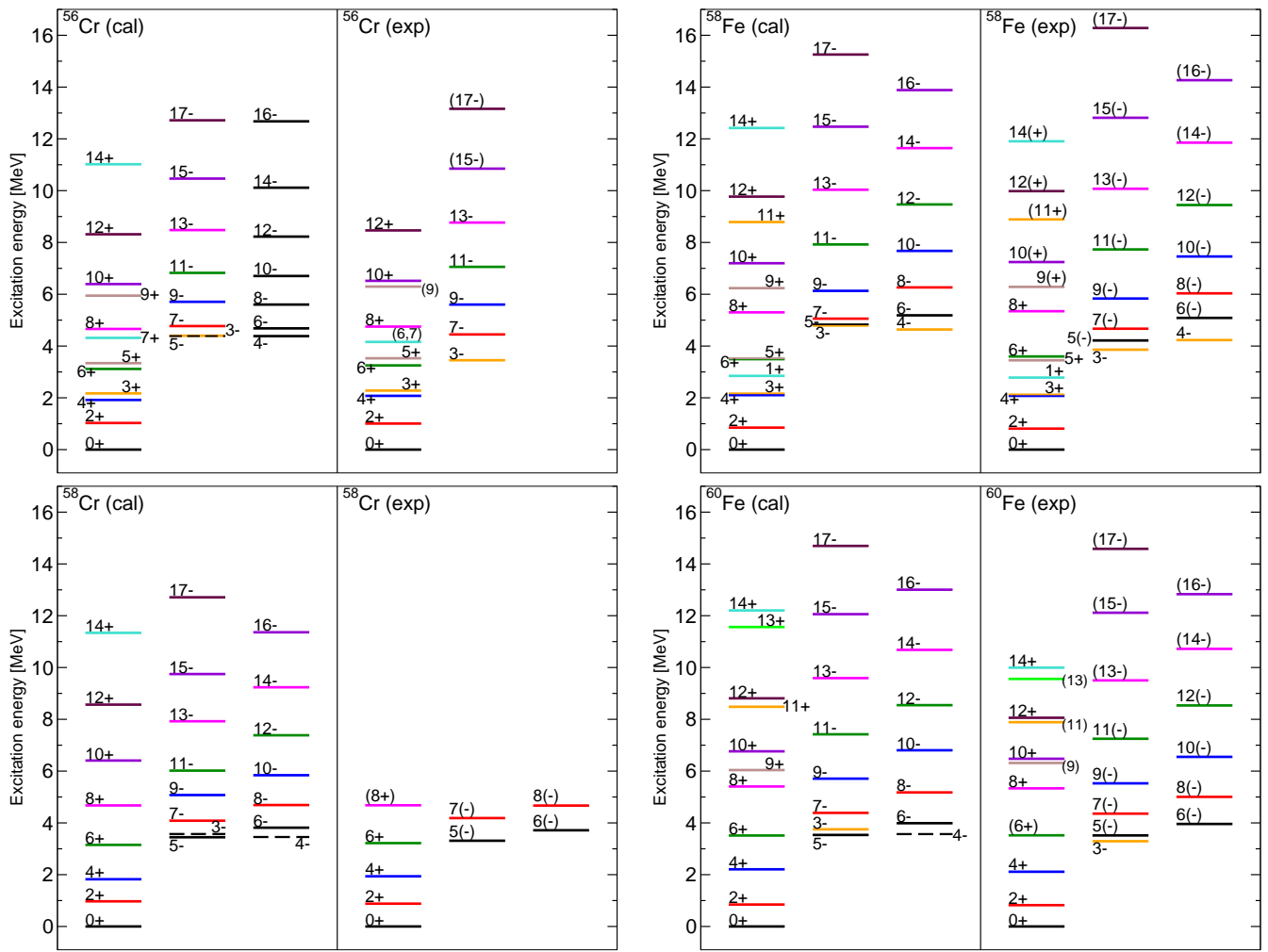

\begin{tabular}{c}
\begin{minipage}{0.5\hsize}
  \begin{center}
  \includegraphics[width=8.6cm]{56Cr_lv.eps}
  \end{center}
\end{minipage}
\begin{minipage}{0.5\hsize}
  \begin{center}
  \includegraphics[width=8.6cm]{58Fe_lv.eps}
  \end{center}
\end{minipage}
\vspace{1mm}
\end{tabular}
\begin{tabular}{c}
\begin{minipage}{0.5\hsize}
  \begin{center}
  \includegraphics[width=8.6cm]{58Cr_lv.eps}
  \end{center}
\end{minipage}
\begin{minipage}{0.5\hsize}
  \begin{center}
  \includegraphics[width=8.6cm]{60Fe_lv.eps}
  \end{center}
\end{minipage}
\end{tabular}
\caption{\label{fig:elv_even}(Color online) Calculated and experimental 
energy levels of the lowest states for each spin and parity 
in even-mass Cr and Fe isotopes. The notation is the same as 
in Fig.~\ref{fig:elv_odd}. The experimental data are taken from 
\cite{fe59-60,cr56_cr58,fe58,NDS_A56,NDS_A58,NDS_A60}. 
In $^{58}$Fe, the experimental negative-parity states, 
except for the $3_{1}^{-}$ and $4_{1}^{-}$ states, are taken from 
Band $2$ and $3$ in \cite{fe58}, where these states are discussed based on 
the projected shell-model calculation \cite{PSM_fe58} interpreting them 
as negative-parity states.}
\end{figure*}
\begin{figure}
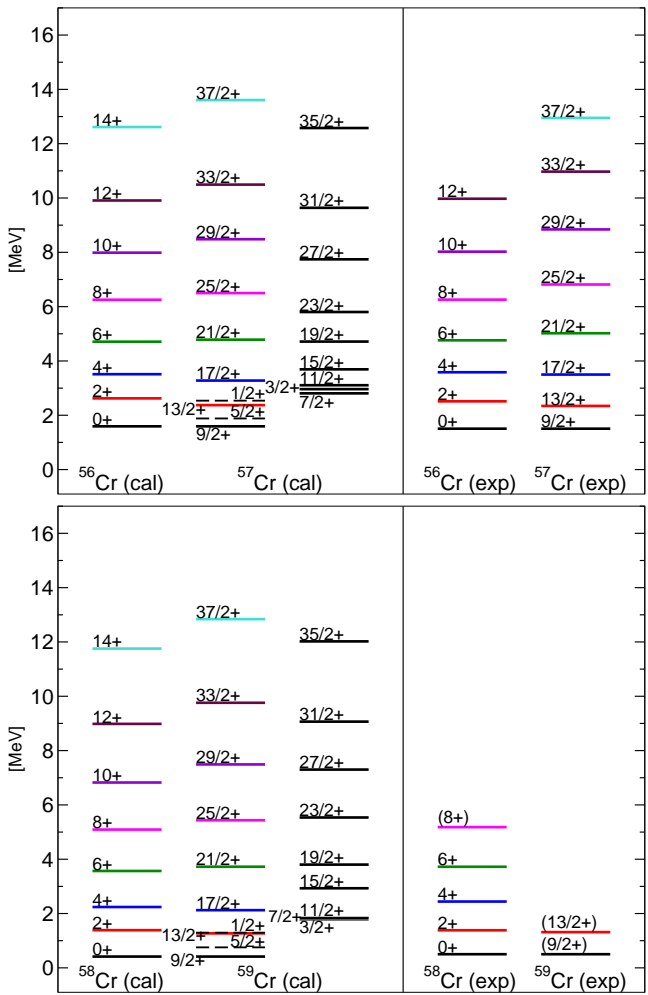

  \begin{tabular}{c}
  \includegraphics[width=8.5cm]{Comp_Cr56-57_lv.eps}\\
  \includegraphics[width=8.5cm]{Comp_Cr58-59_lv.eps}
  \end{tabular}
\caption{\label{fig:comp_cr58-59}(Color online) Comparison between 
the level spacing of the yrast bands in $^{56,58}$Cr and the bands built on 
$9/2_{1}^{+}$ in $^{57,59}$Cr.}
\end{figure}
\begin{figure*}[t]
\begin{tabular}{c|c}
\begin{minipage}{0.5\hsize}
  \begin{center}
  \includegraphics[width=9.2cm]{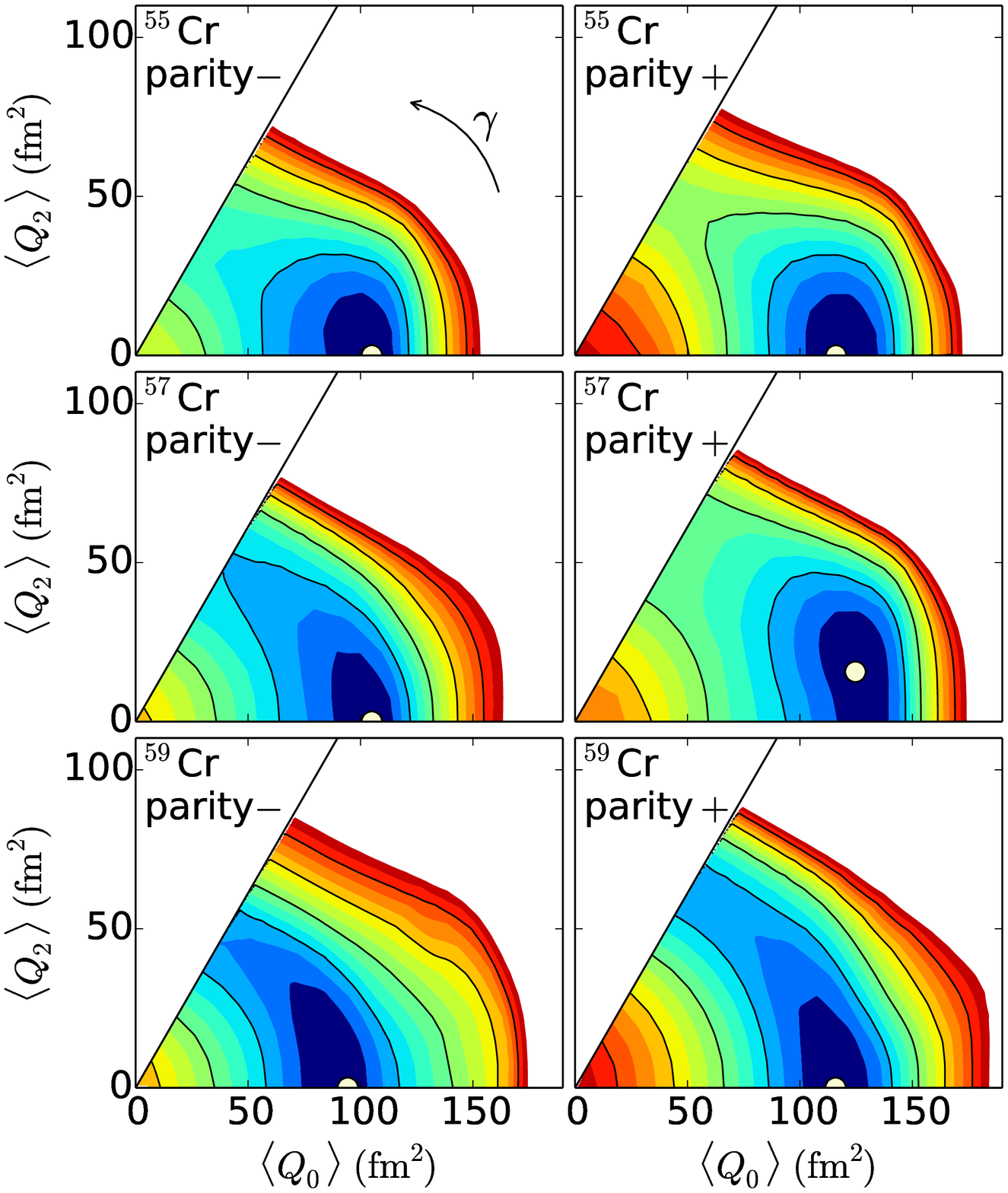}
  \end{center}
\end{minipage}
&
\begin{minipage}{0.5\hsize}
  \begin{center}
  \includegraphics[width=9.2cm]{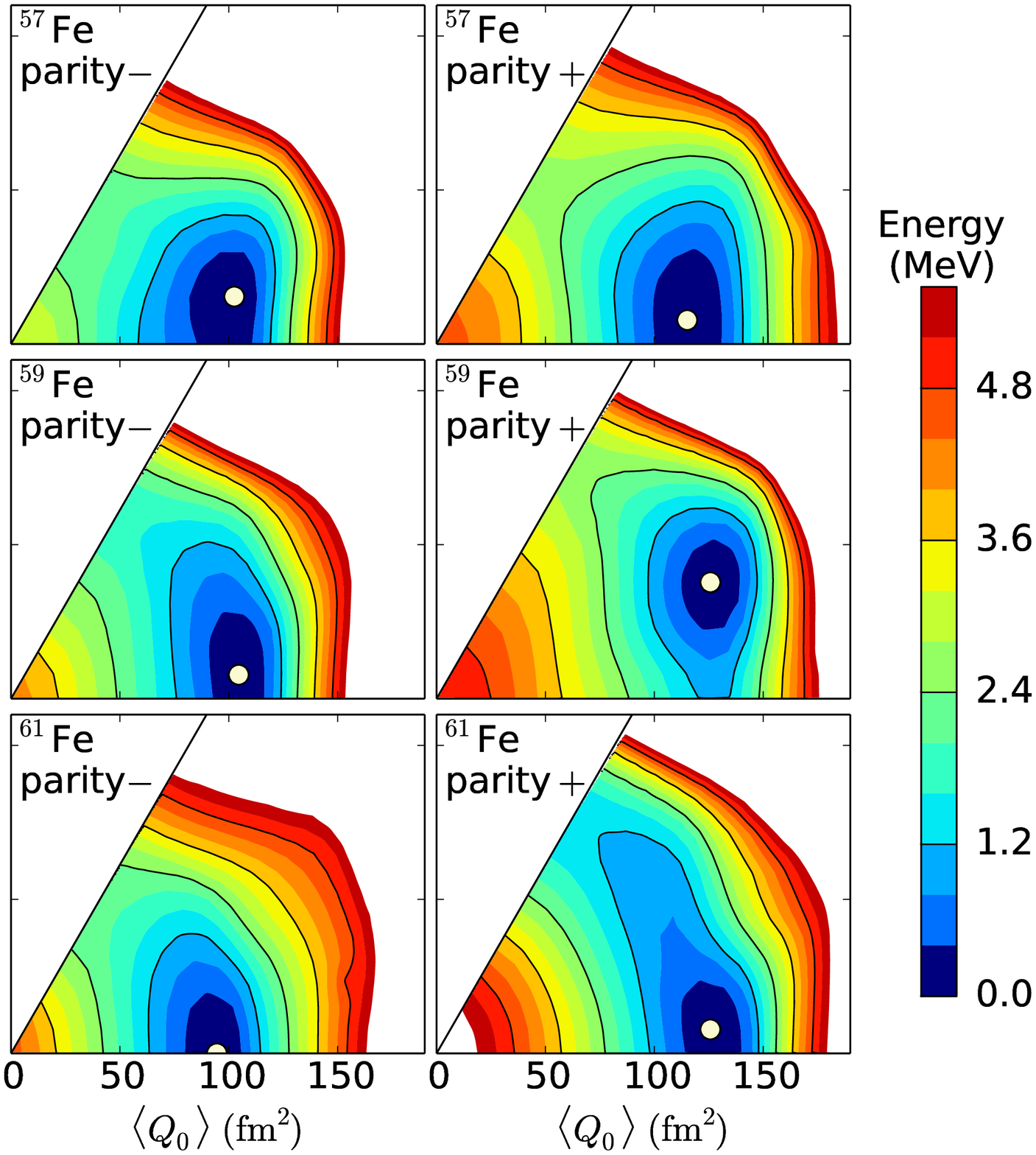}
  \end{center}
\end{minipage}
\end{tabular}
\caption{\label{fig:pes_odd}(Color online) Total energy surfaces for 
the negative(natural)- and positive(unnatural)-parity states 
in odd-mass nuclei. The energy minima are represented by the points of 
open circles. The energy contours are based on the energy minima starting 
from $0$ MeV.}
\end{figure*}

\section{\label{sec:results}Results and Discussion}

\subsection{Energy levels}

The results of the present shell-model calculation with GXPF1Br + 
$V_{\rm MU}$(modified) are compared with 
the experimental data for Cr and Fe isotopes. 
We focus here on the lowest states for a given spin $I$ and parity $\pi=\pm$. 
The results for odd-mass and even-mass nuclei are shown in 
Figs.~\ref{fig:elv_odd} and \ref{fig:elv_even}, respectively, 
where the unnatural-parity states are classified into 
two $\Delta I=2$ sequences.

For the odd-mass $^{55-59}$Cr and $^{57-61}$Fe, 
as seen in Fig.~\ref{fig:elv_odd}, the negative-parity states are 
described well by the shell-model calculation within the $fp$-shell model 
space. Furthermore, almost all of the observed positive-parity states are 
reproduced excellently by the present shell-model calculation. 
For the highest-spin $37/2_{1}^{+}$ state in $^{57}$Cr, 
the backbending indicated by the small energy interval between 
the $37/2_{1}^{+}$ and $33/2_{1}^{+}$ \cite{cr56_cr58} is not well 
reproduced in this calculation. This shows the limitation of description 
within the present model space and truncation. 
The theoretical results show that the $1/2_{1}^{+}$ and $5/2_{1}^{+}$ states 
lie at higher energy than the $9/2_{1}^{+}$ state in each nucleus. 
This means that each $9/2_{1}^{+}$ state is the lowest state in its band.
Actually, the $9/2_{1}^{+}$ and $5/2_{1}^{+}$ states observed in $^{55}$Cr 
and $^{57,59}$Fe are almost degenerate. 
The present shell-model calculation predicts that 
the other positive-parity band including $7/2_{1}^{+}$, 
$11/2_{1}^{+}$, ... is located higher than $13/2_{1}^{+}$ in each nucleus.

For even-mass $^{56,58}$Cr and $^{58,60}$Fe, 
as seen in Fig.~\ref{fig:elv_even}, 
the present calculation excellently describes the positive- and 
negative-parity states except for some of the high-spin states and 
the $3_{1}^{-}$ states. 
The observed $3_{1}^{-}$ states lie at slightly lower 
energies than the calculated ones, which may indicate that these $3_{1}^{-}$ 
states are described not as a single particle-hole configuration 
but as collective states, for instance octupole vibrational states. 
The relative positions of the negative-parity odd-$I$ and even-$I$ bands are 
well reproduced with the present calculation. 
In $^{58}$Fe, for instance, the $6_{1}^{-}$ and $8_{1}^{-}$ levels are 
located higher than the $7_{1}^{-}$ and $9_{1}^{-}$ levels, respectively, 
whereas higher-spin states follow the regular ordering according to spin. 
In contrast, we predict that the $\Delta I=1$ partners such as 
$6_{1}^{-}$ and $7_{1}^{-}$ lie very close up to high-spin states 
for $^{56}$Cr.

One of the interesting features for these Cr and Fe nuclei is that 
the level spacing between $\Delta I=2$ of the band built 
on $9/2_{1}^{+}$ in the odd-mass $A$ nuclei is similar to 
that of the $0_{1}^{+}$ band in the neighboring even-mass $A-1$ nuclei. 
For $^{56-57,58-59}$Cr, in Fig.~\ref{fig:comp_cr58-59}, 
we compare the level spacing of the positive-parity yrast states 
between even- and odd-mass nuclei, which has already been discussed 
for $^{54-55}$Cr \cite{cr55} and $^{56-57,58-59,60-61}$Fe \cite{fe61}. 
One can see the similar level spacing between the yrast band in 
even-mass nuclei and the band of the $9/2_{1}^{+}$ in odd-mass nuclei. 
Although the experimental information for the high-spin states in $^{59}$Cr 
is not enough, the present calculation predicts this similarity of 
the level spacing between $^{58-59}$Cr. 
This level structure indicates the decoupling limit of 
the particle-plus-rotor model \cite{Ring, part_rot} 
as discussed in the next subsection.
\begin{figure*}
\begin{tabular}{c|c}
\begin{minipage}{0.5\hsize}
  \begin{center}
  \includegraphics[width=9.0cm]{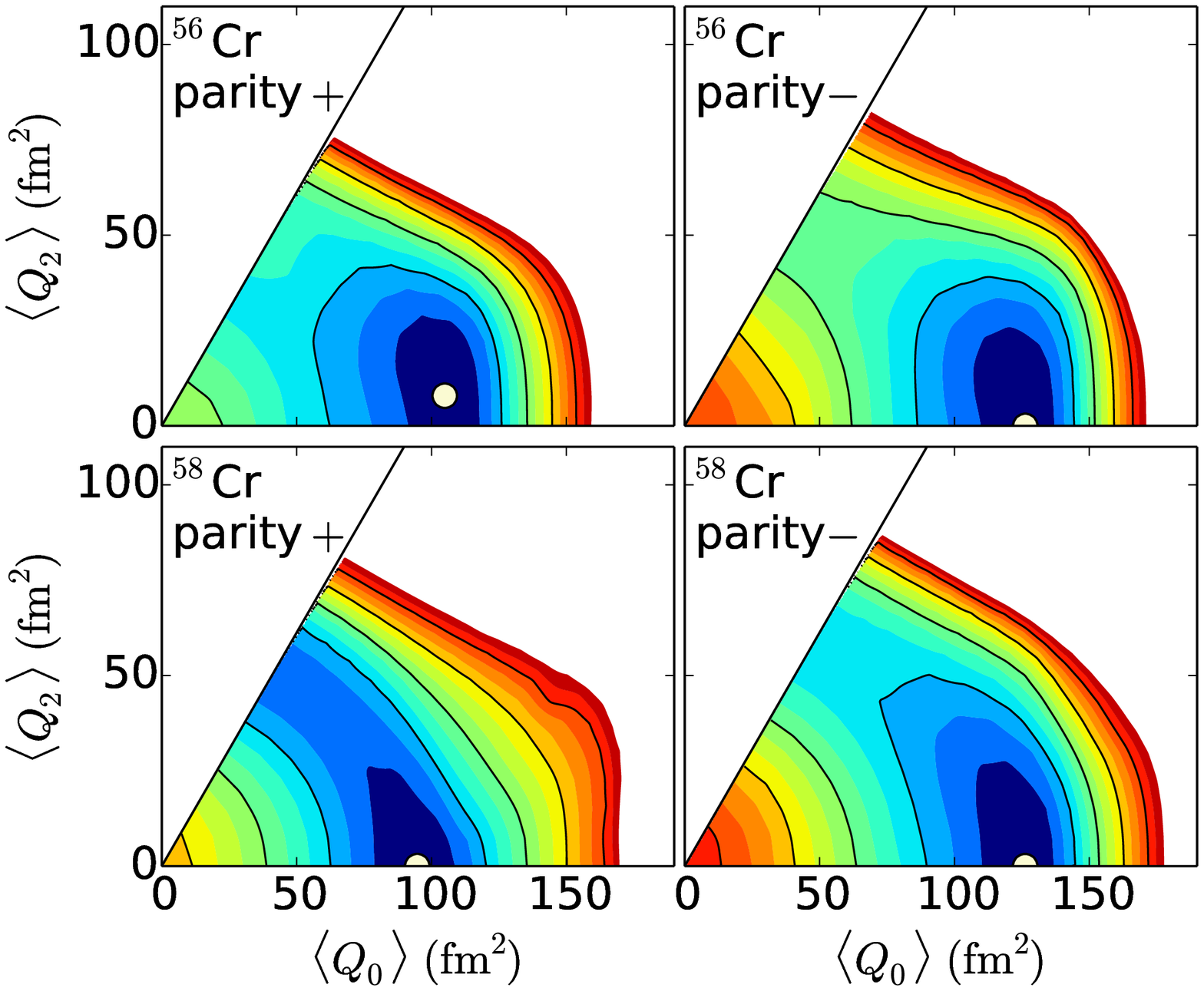}
  \end{center}
\end{minipage}
&
\begin{minipage}{0.5\hsize}
  \begin{center}
  \includegraphics[width=9.0cm]{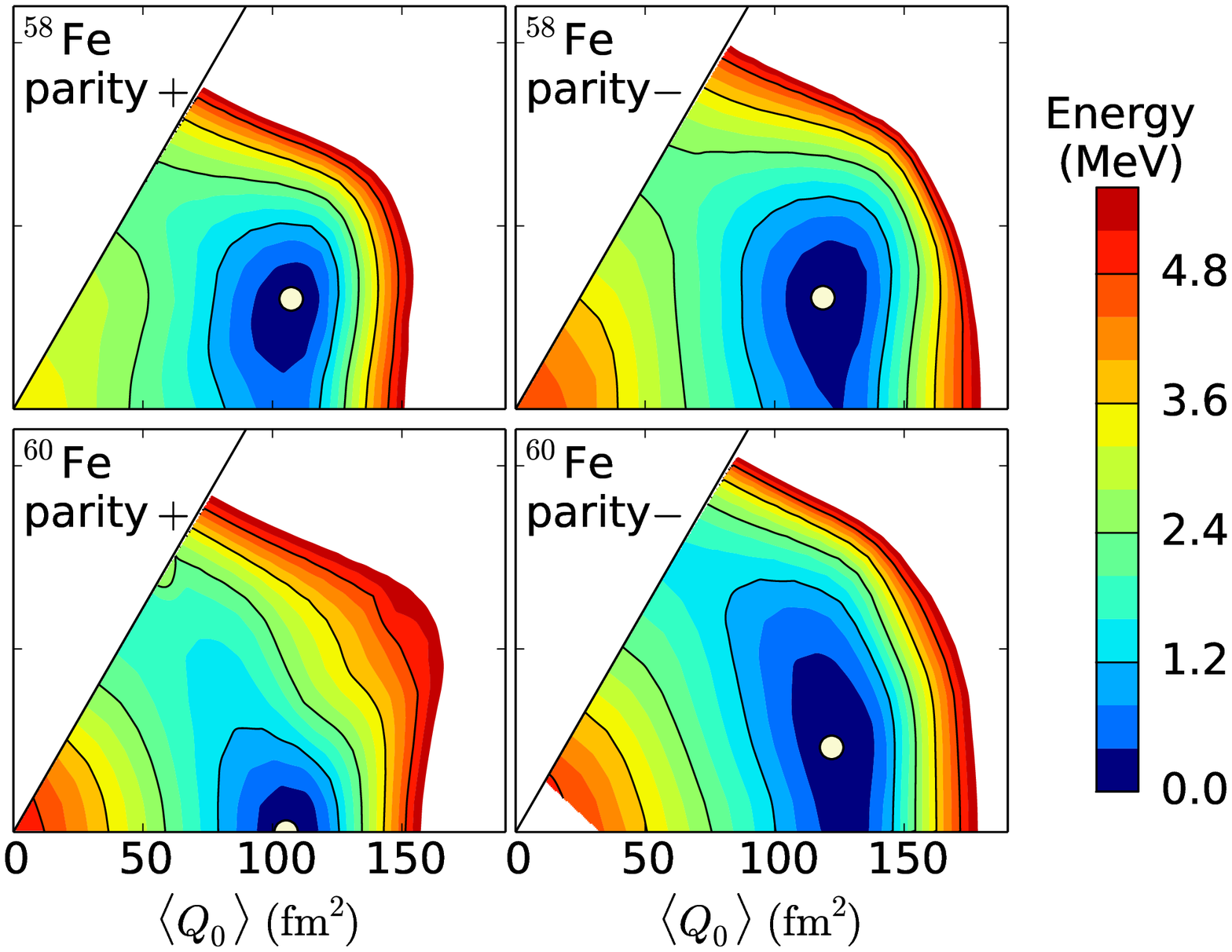}
  \end{center}
\end{minipage}
\end{tabular}
\caption{\label{fig:pes_even}(Color online) The total energy surfaces for 
the positive(natural)- and negative(unnatural)-parity states 
in even-mass nuclei. The notations are the same as in Fig.~\ref{fig:pes_odd}.}
\end{figure*}
\begin{figure*}
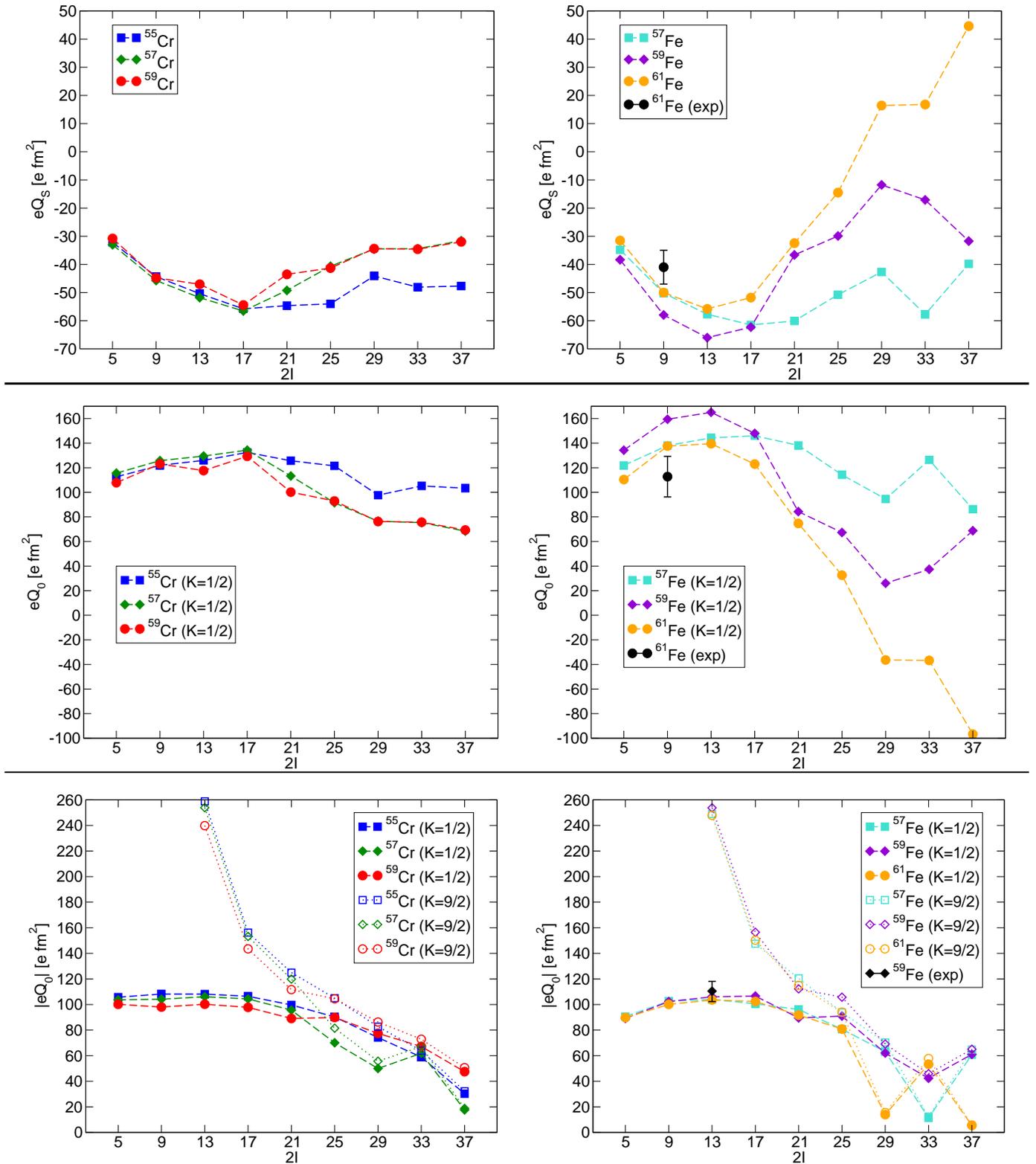

\begin{tabular}{c}
\begin{minipage}{0.5\hsize}
  \begin{center}
  \includegraphics[width=8.4cm]{Qs-Cr_dat.eps}
  \end{center}
\end{minipage}
\begin{minipage}{0.5\hsize}
  \begin{center}
  \includegraphics[width=8.4cm]{Qs-Fe_dat.eps}
  \end{center}
\end{minipage}
\vspace{0.5mm}
\\
\hline\\
\begin{minipage}{0.5\hsize}
  \begin{center}
  \includegraphics[width=8.55cm]{Q0a-Cr_dat.eps}
  \end{center}
\end{minipage}
\begin{minipage}{0.5\hsize}
  \begin{center}
  \includegraphics[width=8.55cm]{Q0a-Fe_dat.eps}
  \end{center}
\end{minipage}
\vspace{0.5mm}
\\
\hline\\
\begin{minipage}{0.5\hsize}
  \begin{center}
  \includegraphics[width=8.4cm]{Q0b-Cr_dat.eps}
  \end{center}
\end{minipage}
\begin{minipage}{0.5\hsize}
  \begin{center}
  \includegraphics[width=8.4cm]{Q0b-Fe_dat.eps}
  \end{center}
\end{minipage}
\end{tabular}
\caption{\label{fig:q_mom}(Color online) The electric spectroscopic and 
intrinsic quadrupole moments for the members of the band built 
on $9/2_{1}^{+}$ in odd-mass Cr and Fe nuclei. 
The horizontal axes denote the spin $I$ values of the states as $2I$. 
The upper panels show the electric spectroscopic 
quadrupole moments $Q_{S}$. The middle and lower panels show 
the intrinsic quadrupole moments $Q_{0}$ solved by Eq.~(\ref{eq:int_q}) 
assuming $K=1/2$ and the absolute values of $Q_{0}$ solved 
by Eq.~(\ref{eq:int_q_be2}), respectively. 
The experimental spectroscopic quadrupole moment $Q_{S}$ of 
$^{61}$Fe \cite{fe61_sqm} labeled ``$^{61}$Fe(exp)'' is assumed to be negative. 
The absolute value of $Q_{0}$ labeled ``$^{59}$Fe(exp)'' is evaluated 
with the experimental $B(E2)$ values \cite{NDS_A59} assuming $K=1/2$.}
\end{figure*}

\subsection{Deformations}

In this subsection, we discuss the deformations for natural- and 
unnatural-parity states in neutron-rich Cr and Fe nuclei in terms of 
their intrinsic quadrupole moments. 
There is no straightforward way to determine the intrinsic quadrupole moments 
with the shell-model calculations because the shell-model wave functions 
are described in the laboratory frame. 
Here we probe deformations from two different approaches. 
One is approximating the wave function in the intrinsic frame. 
The other is extracting intrinsic deformations from shell-model 
wave functions using the geometric model. 

The intrinsic-frame approximation in the shell model is often performed 
with the $Q$-constrained Hartree-Fock method \cite{ni56,MCSM}. 
This method is, however, not suitable for deducing intrinsic wave 
functions for unnatural-parity states which are usually located higher 
than natural-parity states. 
In the present study, we calculate the total energy surfaces 
separately for positive- and negative-parity states 
by performing the $Q$-constrained Hartree-Fock calculation 
with variation after parity projection. 
We utilize the parity-projected wave function 
$|\Phi^{\pi}\rangle$ written as 
\begin{eqnarray}
|\Phi^{\pi}\rangle
=
\frac{1+\pi\Pi}{2}\,|\Phi\rangle,
\quad
|\Phi\rangle
=
\prod_{i}^{N_{f}}
\left(\sum_{l}^{N_{s}} D_{li}\,c^{\dagger}_{l}\right)
|-\rangle,
\label{eq:ppwf}
\end{eqnarray}
where $\Pi$ represents the space-reflection operator, $\pi$ denotes 
the parity, $\pi=\pm$, $N_{f}$ and $N_{s}$ represent the numbers of 
the valence particles and the single-particle orbits in the model space, 
respectively, and $|-\rangle$ denotes an inert core. 
The coefficients $D$ in $|\Phi\rangle$ of Eq.~(\ref{eq:ppwf}) are 
determined to minimize the energy 
$\langle\Phi^{\pi}|H|\Phi^{\pi}\rangle/\langle\Phi^{\pi}|\Phi^{\pi}\rangle$ 
with the constraints of quadrupole moments 
$\langle Q_0 \rangle = \langle\Phi^{\pi}|Q_{0}|\Phi^{\pi}\rangle
/\langle\Phi^{\pi}|\Phi^{\pi}\rangle$ and 
$\langle Q_2\rangle = \langle\Phi^{\pi}|Q_{2}|\Phi^{\pi}\rangle
/\langle\Phi^{\pi}|\Phi^{\pi}\rangle$, 
where the intrinsic axes are chosen to satisfy 
$\langle\Phi^{\pi}|Q_{\pm 1}|\Phi^{\pi}\rangle = 0$. 
$Q_{\mu}$ is the $\mu$ component of the mass quadrupole operator.
Note that the wave function of Eq.~(\ref{eq:ppwf}) is solved by using 
the shell-model Hamiltonian. 

The total energy surfaces for the negative- and 
positive-parity states in odd-mass $^{55-59}$Cr and $^{57-61}$Fe are 
shown in Fig.~\ref{fig:pes_odd}. 
Those energy surfaces indicate the dominance of prolate deformation. 
For the negative-parity states, 
the quadrupole moments at the energy minima are around $100$ fm$^{2}$. 
For the positive-parity states, those values are enlarged to be 
around $120$ fm$^{2}$. 
The wave functions near the energy minima of the positive-parity 
energy surfaces are dominated by the excitation of one neutron into 
the $0g_{9/2}$ orbit. 
This confirms the validity of the truncation of the model space 
introduced in this study. 
While the one-neutron excitation into the $0g_{9/2}$ orbit induces 
larger deformation by $\sim 20$\%, the shape is not changed significantly.
In $^{59}$Cr, the energy surfaces are $\gamma$-soft 
for both negative- and positive-parity states. 
In $^{61}$Fe, $\gamma$-softness is seen for the positive-parity state. 
For the positive-parity state of $^{59}$Fe, triaxiality is well developed.

The situation of the even-mass Cr and Fe nuclei is very similar to 
that of the odd-mass nuclei as seen in Fig.~\ref{fig:pes_even}. 
These energy surfaces indicate the dominance of prolate deformation. 
For the negative-parity states in even-mass nuclei, 
larger prolate deformations are induced by 
the excitation of one neutron into the $\nu0g_{9/2}$ orbit. 
In $^{58}$Fe, triaxiality is developed for both the positive- and 
negative-parity states as suggested by 
the $\gamma$-vibrational band \cite{fe58,fe58_old}. 
In $^{60}$Fe, triaxiality is developed for the negative-parity state.

The intrinsic quadrupole moment $Q_{0}$ can also be deduced from 
the spectroscopic quadrupole moment $Q_{S}$ by assuming 
the $K$ quantum number \cite{Bohr} as 
\begin{eqnarray}
Q_{S} = \frac{3K^{2}-I(I+1)}{(I+1)(2I+3)}Q_{0},
\label{eq:int_q}
\end{eqnarray}
where $I$ denotes the spin of the state. $Q_{S}$ is an observable 
and is also calculated directly by the shell model. 
We calculate $Q_{S}$ with the effective charges $e_{\pi}=1.5e$ and 
$e_{\nu}=0.5e$. The results for the band members starting from 
$9/2_{1}^{+}$ are shown in Fig.~\ref{fig:q_mom}, together with 
the experimental data for the $9/2_{1}^{+}$ state in $^{61}$Fe \cite{fe61_sqm}. 
The calculation reasonably reproduces the measured value. 
It should be noted that the magnetic moment of this state $-1.031(9)\mu_{N}$
\cite{fe61_g} is also close to the calculated value $-0.934\mu_{N}$ 
using the free-nucleon $g$ factors. 

Since $Q_{0}$ is identical among the members of an ideal rotor, 
the stability of $Q_{0}$ within a band 
provides a key to determining the intrinsic state. 
Here, we take the two possibilities of $K$ for the band members built on 
$9/2_{1}^{+}$: $K=1/2$ and $9/2$ correspond to prolate and oblate shapes, 
respectively. 
Figure~\ref{fig:q_mom} shows the intrinsic quadrupole moment $Q_{0}$ 
assuming $K=1/2$ for odd-mass nuclei $^{55-59}$Cr and $^{57-61}$Fe. 
As seen in Fig.~\ref{fig:q_mom}, the $Q_{0}$ value is nearly constant in Cr 
isotopes, whereas it decreases for high-spin states in Fe isotopes. 
This decrease might be due to change in shape or $K$ number, 
as suggested by the predicted backbending in the $29/2_{1}^{+}$ 
of $^{61}$Fe (see Fig.~\ref{fig:elv_odd}). 
If $K=9/2$ is assumed, on the other hand, 
we find a singular behavior of the $Q_{0}$ value for each nucleus: 
the $Q_{0}$ value is about $-600$ $e$ fm$^{2}$ in the $13/2_{1}^{+}$ state 
and it suddenly transits into the large positive value of 
about $600$ $e$ fm$^{2}$ in the $17/2_{1}^{+}$ state. 

The intrinsic quadrupole moment $Q_{0}$ is also connected to the  
$B(E2,I \rightarrow I-2)$ value \cite{Bohr}:
\begin{eqnarray}
B(E2,I \rightarrow I-2) = \frac{5}{16\pi}
\left( \langle IK20|I-2\;K \rangle \right)^{2} Q_{0}^{2},
\label{eq:int_q_be2}
\end{eqnarray}
where $\langle IK20|I-2\;K \rangle$ denotes 
the Clebsch-Gordan coefficient. Note that the sign of $Q_{0}$ cannot be 
determined using Eq.~(\ref{eq:int_q_be2}). 
We calculate the $B(E2)$ values with the same effective charges 
as the ones used for $Q_{S}$. 
The lowest panels of Fig.~\ref{fig:q_mom} show the absolute values of $Q_{0}$ 
solved by Eq.~(\ref{eq:int_q_be2}) 
under the assumptions of $K=1/2$ and $K=9/2$.
The present calculation reproduces the experimental 
$B(E2,13/2_{1}^{+} \rightarrow 9/2_{1}^{+})$ in $^{59}$Fe. 
Similar to $Q_S$, the $B(E2)$ values lead to much more stable 
intrinsic quadrupole moments with $K=1/2$ than with $K=9/2$. 
Although the mixing of the $K$ quantum numbers occurs in reality 
due to the Coriolis effect, 
the present calculation supports the dominance of $K=1/2$ 
rather than $K=9/2$ and thus 
prolate deformation for the odd-mass Cr and Fe nuclei studied. 

We point out here that the overall prolate deformation 
in the positive-parity states 
of $^{55,57,59}$Cr and $^{57,59,61}$Fe is indeed suggested 
by the measured energy levels, 
according to the analysis of the particle-plus-rotor model 
presented in \cite{Ring, part_rot}. 
It is quite reasonable to apply 
the particle-plus-rotor model in which the last neutron occupies 
the $j=9/2$ orbit because the $9/2_{1}^{+}$ states of those nuclei 
have rather large $\nu 0g_{9/2}$ spectroscopic factors as presented 
in the next section. 
The level structure realized by the particle-plus-rotor model 
strongly depends on the strength of the Coriolis term relative to 
the $\Omega^{2}$ term (originating from the Nilsson levels and recoil term), 
where $\Omega$ is the projection of the angular momentum $j$ of an orbit onto 
the symmetry axis. We here consider the case where the $j$ orbit is occupied 
by the last nucleon alone. 
For oblate deformation having the maximum $\Omega$ ($=j$), 
the Coriolis term is less dominant 
because its matrix element is proportional to 
$(j(j+1)-\Omega^{2})^{1/2}$. 
The energy levels then follow the strong-coupling limit of $I(I+1)/2\cal{I}$ 
($I=\Omega, \Omega+1, \Omega+2, \ldots$), where 
$\cal{I}$ is the moment of inertia. 
For prolate deformation, on the other hand, the Coriolis term can 
dominate over the $\Omega^{2}$ term especially when a high-$j$ orbit 
is involved. The energy levels in this case follow 
the decoupling limit of $(I-\alpha)(I-\alpha+1)/2\cal{I}$, 
where $\alpha$ is the projection of $j$ onto the rotation axis. 
The lowest-lying states take the largest possible $\alpha$. 
Since symmetry considerations require $I-\alpha$ to be even, 
favored states with $\alpha = j$ consist of 
$I=j, j+2, j+4, \ldots$ and unfavored states with 
$\alpha = j-1$ consist of $I=j+1, j+3, j+5, \ldots$. 
As a result, two major differences arise between prolate and oblate shapes. 
One is that the splitting of the favored and unfavored states is seen only 
for prolate deformation. The other is that energy spacings in 
the $I=j, j+2, j+4, \ldots$ levels are identical with those of even-even core, 
$0^{+}, 2^{+}, 4^{+}, \ldots$, for prolate deformation 
but are much wider for oblate deformation. 
Those differences are also obtained from a microscopic calculation 
using the projected shell model \cite{cr59_def}. 
The observed levels shown in Figs.~\ref{fig:elv_odd} and \ref{fig:comp_cr58-59} 
clearly support the decoupling limit which is realized for prolate deformation. 
Oblate deformation is quite unlikely but 
might not be completely excluded if deformation is small. 
Measuring unfavored states will provide additional information on shape. 
\begin{figure}[h]
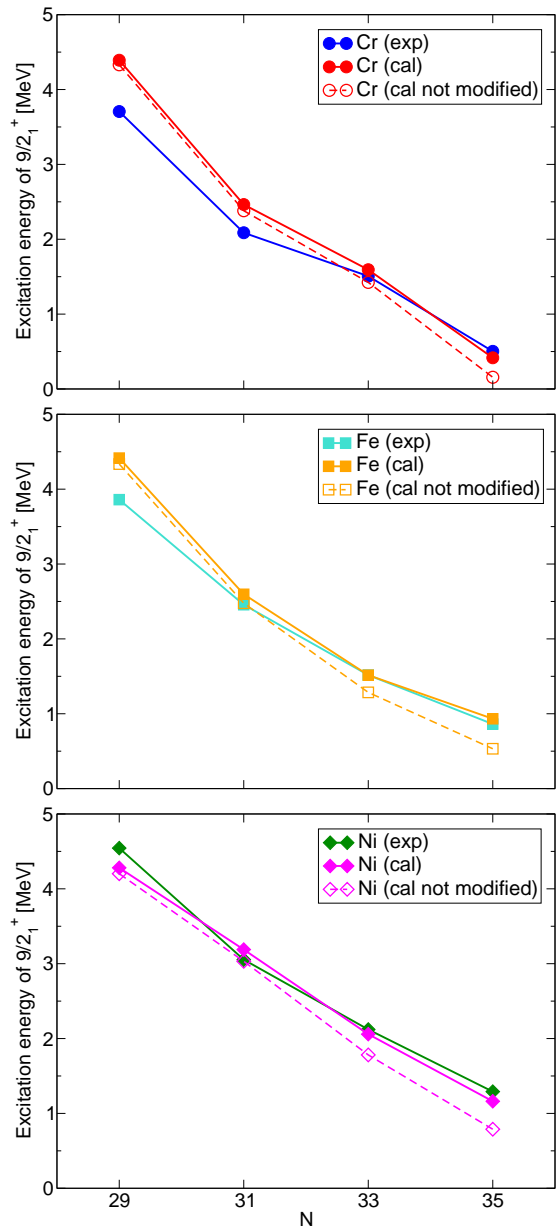

\begin{tabular}{c}
  \includegraphics[width=7.25cm]{9_2_EX_Cr.eps}\\
  \includegraphics[width=7.25cm]{9_2_EX_Fe.eps}\\
  \includegraphics[width=7.25cm]{9_2_EX_Ni.eps}
\end{tabular}
\caption{\label{fig:9_2_EX}(Color online) The excitation energy of 
$9/2_{1}^{+}$ in odd-mass Cr, Fe and Ni isotopes from $N=29$ to $N=35$. 
The plots labeled ``(cal)'' and ``(exp)'' represent the present results 
and the experimental data, respectively. 
The plots labeled ``(cal not modified)'' show the results without 
the modification of the two-body matrix elements described in 
Sec.~\ref{sec:effh}. 
The experimental data are taken from 
\cite{cr55,cr57,cr59,fe59-60,fe61,NDS_A53,NDS_A55,NDS_A57,NDS_A59,NDS_A61,NDS_A63}. 
In $^{55}$Fe, the state of $3.86$ MeV excitation energy in \cite{NDS_A55} 
is adopted as the experimental $9/2^{+}$ state because it has the largest 
spectroscopic factor of $9/2^{+}$ in the $(d,p)$ reaction.}
\end{figure}

\section{\label{sec:n0g9}Evolution of the $\nu 0g_{9/2}$ orbit}

In this section, we examine what causes the sharp drop of 
the $9/2_{1}^{+}$ level with increasing neutron number 
observed in Cr and Fe isotopes. A particular attention is paid to 
the relation to the evolution of the $\nu 0g_{9/2}$ orbit. 
For this purpose, we also investigate the $9/2_{1}^{+}$ levels in 
Ni isotopes, which should more directly reflect spherical 
single-particle structure. It is noted that the maximum $M$-scheme 
dimension reaches $1.8 \times 10^{10}$ for $^{59}$Ni.

We first confirm the descriptive power of our approach. 
Figure~\ref{fig:9_2_EX} shows the results of the excitation energy of 
$9/2_{1}^{+}$ from $N=29$ to $N=35$. The sharp drop of $9/2_{1}^{+}$ 
is reproduced quite well in the present calculations. 
Although the evolution of the $9/2_{1}^{+}$ level is improved, 
as well as the high-spin levels, by the modification of the Hamiltonian 
introduced in Sec.~\ref{sec:effh}, 
its basic trend is already seen without the modification. 
Thus, the $V_{\rm MU}$ interaction gives a reasonable evolution 
of the $\nu 0g_{9/2}$ orbit, if the $9/2^+_1$ levels are of 
single-particle origin. 
Besides Cr and Fe isotopes, the present calculation also reproduces 
the systematics of $9/2_{1}^{+}$ in Ni isotopes. 
This shows wide applicability of the present Hamiltonian in this mass region.
\begin{table}
\caption{\label{tab:sfac}
Spectroscopic factors of $\nu 0g_{9/2}$ and $\nu 1d_{5/2}$ particle states 
for $9/2_{1}^{+}$ and $5/2_{1}^{+}$, respectively.
The experimental values labeled ``exp'' are taken from the data of
the $(d,p)$ reaction \cite{53Cr_dp,53Cr55Cr_dp,53Cr55Fe_dp,55Cr_dp,55Fe63Ni_dp,55Fe_dp,57Fe_dp1,57Fe_dp2,59Fe_dp,55Cr59Fe59Ni_dp,59Ni_dp1,59Ni_dp2,59Ni_dp3,61Ni_dp1,61Ni_dp2,61Ni_dp3,61Ni_dp4,63Ni_dp1,63Ni_dp2,63Ni_dp3},
the $(\alpha,^{3}\mbox{He})$ reaction \cite{55Fe59Ni_He43}, 
the $(^{7}\mbox{Li},^{6}\mbox{Li})$ reaction \cite{55Fe_Li76}, 
and the $(^{14}\mbox{C},^{13}\mbox{C})$ reaction \cite{59Ni_C1413}. 
Experimental excitation energies of $5/2^{+}$ in $^{53}$Cr and $^{55}$Fe
are $4.13$ MeV and $4.46$ MeV, respectively.
Calculated excitation energies of $5/2_{1}^{+}$ are $5.09$ MeV and
$5.19$ MeV for $^{53}$Cr and $^{55}$Fe, respectively.}
\begin{ruledtabular}
\begin{tabular}{ccc}
$C^{2}S$($9/2_{1}^{+}$)&cal&exp\\
\hline
$^{53}$Cr&0.564&0.520~\cite{53Cr_dp}, 0.95~\cite{53Cr55Cr_dp}, 0.57~\cite{53Cr55Fe_dp}\\
$^{55}$Cr&0.458&0.67~\cite{53Cr55Cr_dp}, 0.582~\cite{55Cr_dp}\\
$^{57}$Cr&0.479&-\\
$^{59}$Cr&0.498&-\\
\hline
$^{55}$Fe&0.568&0.74~\cite{55Fe63Ni_dp}, 0.465~\cite{55Fe_dp}, 0.375~\cite{55Fe59Ni_He43},\\
         &     &0.67~\cite{55Fe_Li76}\\
$^{57}$Fe&0.494&0.270~\cite{57Fe_dp1}, 0.447~\cite{57Fe_dp2}\\
$^{59}$Fe&0.442&0.510~\cite{59Fe_dp}, 0.38~\cite{55Cr59Fe59Ni_dp}\\
$^{61}$Fe&0.527&-\\
\hline
$^{57}$Ni&0.611&-\\
$^{59}$Ni&0.580&0.84~\cite{59Ni_dp1}, 0.47~\cite{59Ni_dp2}, 0.56~\cite{55Cr59Fe59Ni_dp},\\
         &     &0.381~\cite{59Ni_dp3}, 0.390~\cite{55Fe59Ni_He43}, 0.69~\cite{59Ni_C1413}\\
$^{61}$Ni&0.503&0.62~\cite{61Ni_dp1}, 0.750~\cite{61Ni_dp2}, 0.8450~\cite{61Ni_dp3},\\
         &     &0.537~\cite{61Ni_dp4}\\
$^{63}$Ni&0.446&0.61~\cite{55Fe63Ni_dp}, 0.672~\cite{63Ni_dp1}, 0.75~\cite{63Ni_dp2},\\
         &     &0.75~\cite{63Ni_dp3}\\
\hline
\hline
$C^{2}S$($5/2_{1}^{+}$)&cal&exp\\
\hline
$^{53}$Cr&0.161&0.090~\cite{53Cr_dp}, 0.13~\cite{53Cr55Cr_dp}, 0.10~\cite{53Cr55Fe_dp}\\
$^{55}$Cr&0.142&0.225~\cite{53Cr55Cr_dp}, 0.145~\cite{55Cr_dp}, 0.17~\cite{55Cr59Fe59Ni_dp}\\
$^{57}$Cr&0.148&-\\
$^{59}$Cr&0.152&-\\
\hline
$^{55}$Fe&0.117&0.172~\cite{55Fe63Ni_dp}, 0.079~\cite{55Fe_dp}, 0.13~\cite{53Cr55Fe_dp},\\
         &     &0.25~\cite{55Fe_Li76}\\
$^{57}$Fe&0.120&0.110~\cite{57Fe_dp1}, 0.114~\cite{57Fe_dp2}\\
$^{59}$Fe&0.113&0.128~\cite{59Fe_dp}, 0.11~\cite{55Cr59Fe59Ni_dp}\\
$^{61}$Fe&0.133&-\\
\end{tabular}
\end{ruledtabular}
\end{table}

The single-particle properties of the odd-mass Cr, Fe and Ni isotopes 
are probed from the spectroscopic factors for neutron transfer reactions. 
The measured spectroscopic factors are compared to the calculated values 
in Table~\ref{tab:sfac}. 
While the experimental values scatter in some nuclei, the present 
calculation well reproduces their systematic behavior: $\sim 0.5$ 
for the $9/2^+_1$ states and $0.1$-$0.2$ for the $5/2^+_1$ states. 
This indicates that these $9/2_{1}^{+}$ states are the single-particle-like 
states but that the $5/2_{1}^{+}$ states are not. 
Hence, the excitation energy of $9/2_{1}^{+}$ in these nuclei is 
influenced directly by the location of the $\nu 0g_{9/2}$ orbit. 
\begin{figure}
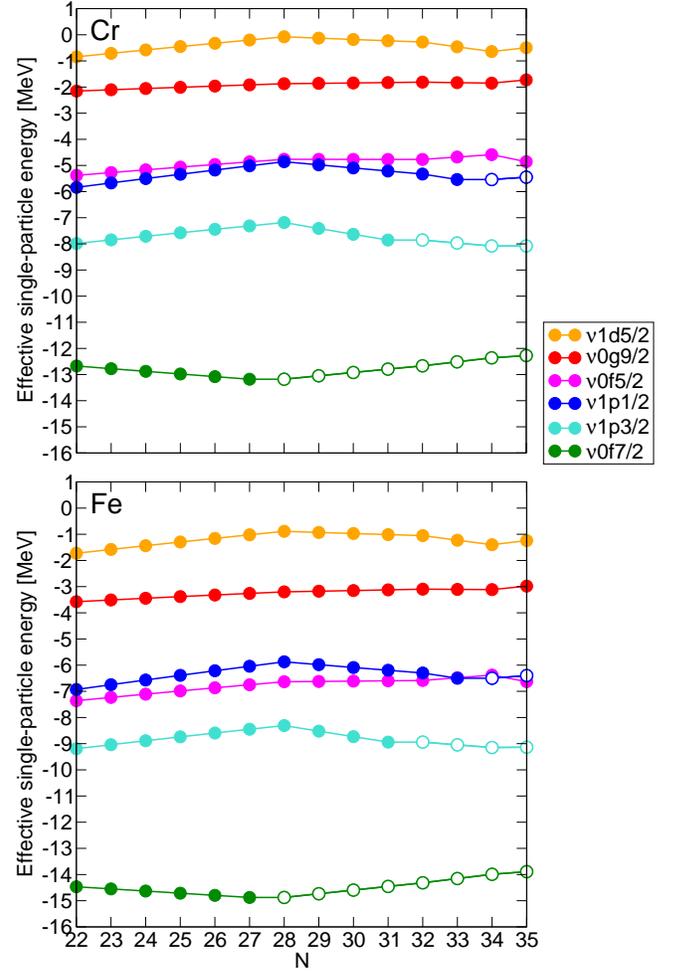

\begin{tabular}{c}
  \includegraphics[width=8.5cm]{Cr-ESPE.eps}\\
  \hspace{-16.5mm}
  \includegraphics[width=6.97cm]{Fe-ESPE.eps}
\end{tabular}
\caption{\label{fig:espe}(Color online) The effective single-particle energies 
(ESPEs) of a neutron for Cr and Fe isotopes. The open circles represent 
the ESPEs calculated as the hole states to specify the occupied states.}
\end{figure}

Neutron effective single-particle energies (ESPEs)~\cite{ESPE} 
in Cr and Fe isotopes are displayed in Fig.~\ref{fig:espe} 
using the present Hamiltonian. 
The ESPEs of $\nu 0g_{9/2}$ are rather constant with increasing neutron number, 
and so are the shell gaps between the $fp$ shell and $\nu 0g_{9/2}$. 
This indicates that the evolution of the $9/2_{1}^{+}$ level 
is explained by the Fermi surface approaching 
the $\nu 0g_{9/2}$ orbit with increasing neutron number. 
This is in contrast to the result of a previous shell-model study 
\cite{Kaneko} in which this evolution occurs due to the lowering 
of the $0g_{9/2}$ orbit. 
This lowering was realized in \cite{Kaneko} by shifting 
the $T=1$ monopole interaction between $0p_{3/2}$ and $0g_{9/2}$ by $-1.0$~MeV. 
It should be noted that the $T=1$ $j\ne j'$ monopole matrix elements 
in realistic interactions are in general rather small \cite{vmu}. 

\begin{figure}
  \includegraphics[width=8.5cm]{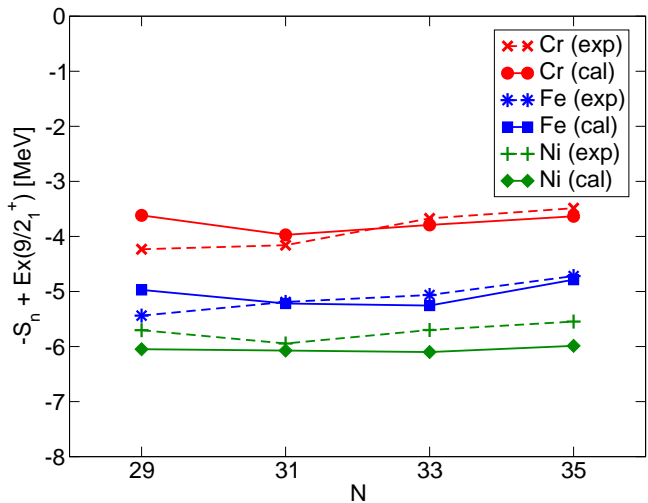}
\caption{\label{fig:sn_ex9}(Color online) Plots of $-S_{n}+E_{x}(9/2_{1}^{+})$. 
Experimental one-neutron separation energies $S_{n}$ are 
taken from \cite{NDS_A53,NDS_A55,NDS_A57,NDS_A59,NDS_A61,NDS_A63}. 
Coulomb energy is corrected in the calculated $S_{n}$ 
according to \cite{gxpf1}.}
\end{figure}

The evolution of the $\nu 0g_{9/2}$ orbit can be probed from experimental 
energies. Namely, in the independent-particle limit, 
the $\nu 0g_{9/2}$ single-particle energy is identical with 
the energy of the $9/2_{1}^{+}$ state measured from 
the adjacent even-even core nucleus, $-S_{n}+E_{x}(9/2_{1}^{+})$. 
This idea can be extended to deformed single-particle energies. 
In Fig.~\ref{fig:sn_ex9}, the $-S_{n}+E_{x}(9/2_{1}^{+})$ values are 
compared between the experimental data and the present calculation. 
The observed stabilities in $-S_{n}+E_{x}(9/2_{1}^{+})$ along 
the Cr, Fe and Ni isotopic chains suggest that 
the $\nu 0g_{9/2}$ single-particle energies are kept nearly constant 
as shown in Fig.~\ref{fig:espe}. 

\section{\label{sec:conclusion}Conclusions}

We have investigated unnatural-parity high-spin states in neutron-rich 
Cr and Fe isotopes by large-scale shell-model calculations 
for the model space of $fp$-shell + $0g_{9/2}$ + $1d_{5/2}$ orbits 
with the truncation allowing $1\hbar\omega$ excitation of a neutron. 
The effective Hamiltonian is composed of GXPF1Br for $fp$-shell orbits 
and a modified $V_{\rm MU}$ for the other parts. 
The shell-model calculations with the present Hamiltonian can describe and 
predict the energy levels of both natural- and unnatural-parity states 
up to the high-spin states in Cr and Fe isotopes with $N\le35$. 
This shell-model calculation has shown that the $9/2_{1}^{+}$ states 
are the lowest positive-parity states in the odd-mass nuclei 
and the level spacings of the bands built on $9/2_{1}^{+}$ 
in the odd-mass $A$ nuclei are similar to those of the $0_{1}^{+}$ bands in 
the neighboring even-mass $A-1$ nuclei.

We have also discussed the deformations of Cr and Fe isotopes 
by utilizing the $Q$-constrained Hartree-Fock calculation with variation after 
parity projection. This study has indicated that the present Cr and Fe nuclei 
have the prolate deformations on the whole. 
In these energy surfaces, the excitation of one neutron into the $\nu0g_{9/2}$ 
orbit plays the role of enhancing the prolate deformation. 
The electric intrinsic quadrupole moments $Q_{0}$, 
which are calculated by using the electric spectroscopic quadrupole 
moments $Q_{S}$ and the $B(E2)$ values obtained in 
the shell-model calculation, have supported the prolate deformations 
of the states in the band built on $9/2_{1}^{+}$ in odd-mass Cr and Fe nuclei 
and have indicated that their dominant $K$ quantum number is $1/2$. 
This situation corresponds to the decoupling limit of the particle-plus-rotor 
model, which can explain the systematics of the positive-parity level schemes 
in these Cr and Fe nuclei.

In the present calculation, the effective single-particle 
energies of $\nu 0g_{9/2}$ in Cr and Fe isotopes are rather constant 
in the region with $N\le35$, which means very weak attraction 
between the $fp$-shell orbits and $0g_{9/2}$ orbit 
as given by $V_{\rm MU}$. 
We have shown that this can be confirmed from the one-neutron separation 
energy ($S_{n}$) and the excitation energy of $9/2_{1}^{+}$ 
($E_{x}(9/2_{1}^{+})$). 
This result indicates that the sharp drop of the $9/2_{1}^{+}$ levels 
in the Cr and Fe nuclei in this mass region 
is explained by the Fermi surface approaching the $\nu0g_{9/2}$ orbit 
with the increase of neutron number.
\vspace{5mm}

\begin{acknowledgments}
This work was supported in part by MEXT SPIRE Field 5 
``The origin of matter and the universe'' and JICFuS.  
This research partly used the computational resources of the K computer 
provided by the RIKEN Advanced Institute for Computational Science through 
the HPCI System Research Project (Project ID:hp130024, hp140210), 
the FX10 supercomputer at the Information Technology Center at the University 
of Tokyo, and the computational resources of the RIKEN-CNS joint research 
project on large-scale nuclear-structure calculations.
\end{acknowledgments}

\nocite{*}

\bibliography{apssamp}

\providecommand{\noopsort}[1]{}\providecommand{\singleletter}[1]{#1}%
\begin{thebibliography}{88}%
\makeatletter
\providecommand \@ifxundefined [1]{%
 \@ifx{#1\undefined}
}%
\providecommand \@ifnum [1]{%
 \ifnum #1\expandafter \@firstoftwo
 \else \expandafter \@secondoftwo
 \fi
}%
\providecommand \@ifx [1]{%
 \ifx #1\expandafter \@firstoftwo
 \else \expandafter \@secondoftwo
 \fi
}%
\providecommand \natexlab [1]{#1}%
\providecommand \enquote  [1]{``#1''}%
\providecommand \bibnamefont  [1]{#1}%
\providecommand \bibfnamefont [1]{#1}%
\providecommand \citenamefont [1]{#1}%
\providecommand \href@noop [0]{\@secondoftwo}%
\providecommand \href [0]{\begingroup \@sanitize@url \@href}%
\providecommand \@href[1]{\@@startlink{#1}\@@href}%
\providecommand \@@href[1]{\endgroup#1\@@endlink}%
\providecommand \@sanitize@url [0]{\catcode `\\12\catcode `\$12\catcode
  `\&12\catcode `\#12\catcode `\^12\catcode `\_12\catcode `\%12\relax}%
\providecommand \@@startlink[1]{}%
\providecommand \@@endlink[0]{}%
\providecommand \url  [0]{\begingroup\@sanitize@url \@url }%
\providecommand \@url [1]{\endgroup\@href {#1}{\urlprefix }}%
\providecommand \urlprefix  [0]{URL }%
\providecommand \Eprint [0]{\href }%
\providecommand \doibase [0]{http://dx.doi.org/}%
\providecommand \selectlanguage [0]{\@gobble}%
\providecommand \bibinfo  [0]{\@secondoftwo}%
\providecommand \bibfield  [0]{\@secondoftwo}%
\providecommand \translation [1]{[#1]}%
\providecommand \BibitemOpen [0]{}%
\providecommand \bibitemStop [0]{}%
\providecommand \bibitemNoStop [0]{.\EOS\space}%
\providecommand \EOS [0]{\spacefactor3000\relax}%
\providecommand \BibitemShut  [1]{\csname bibitem#1\endcsname}%
\let\auto@bib@innerbib\@empty
\bibitem [{\citenamefont {{Huck}{\it~et~al.}}(1985)}]{ca52_1}%
  \BibitemOpen
  \bibfield  {author} {\bibinfo {author} {\bibfnamefont {A.}~\bibnamefont
  {{Huck}{\it~et~al.}}},\ }\href@noop {} {\bibfield  {journal} {\bibinfo
  {journal} {Phys.\ Rev.\ C}\ }\textbf {\bibinfo {volume} {31}},\ \bibinfo
  {pages} {2226} (\bibinfo {year} {1985})}\BibitemShut {NoStop}%
\bibitem [{\citenamefont {{Gade}{\it~et~al.}}(2006)}]{ca52_2}%
  \BibitemOpen
  \bibfield  {author} {\bibinfo {author} {\bibfnamefont {A.}~\bibnamefont
  {{Gade}{\it~et~al.}}},\ }\href@noop {} {\bibfield  {journal} {\bibinfo
  {journal} {Phys.\ Rev.\ C}\ }\textbf {\bibinfo {volume} {74}},\ \bibinfo
  {pages} {021302(R)} (\bibinfo {year} {2006})}\BibitemShut {NoStop}%
\bibitem [{\citenamefont {{Janssens}{\it~et~al.}}(2002)}]{ti54_1}%
  \BibitemOpen
  \bibfield  {author} {\bibinfo {author} {\bibfnamefont {R.~V.~F.}\
  \bibnamefont {{Janssens}{\it~et~al.}}},\ }\href@noop {} {\bibfield  {journal}
  {\bibinfo  {journal} {Phys.\ Lett.\ B}\ }\textbf {\bibinfo {volume} {546}},\
  \bibinfo {pages} {55} (\bibinfo {year} {2002})}\BibitemShut {NoStop}%
\bibitem [{\citenamefont {{Dinca}{\it~et~al.}}(2005)}]{ti54_2}%
  \BibitemOpen
  \bibfield  {author} {\bibinfo {author} {\bibfnamefont {D.-C.}\ \bibnamefont
  {{Dinca}{\it~et~al.}}},\ }\href@noop {} {\bibfield  {journal} {\bibinfo
  {journal} {Phys.\ Rev.\ C}\ }\textbf {\bibinfo {volume} {71}},\ \bibinfo
  {pages} {041302(R)} (\bibinfo {year} {2005})}\BibitemShut {NoStop}%
\bibitem [{\citenamefont {{Chapman}{\it~et~al.}}(1968)}]{cr56_1}%
  \BibitemOpen
  \bibfield  {author} {\bibinfo {author} {\bibfnamefont {R.}~\bibnamefont
  {{Chapman}{\it~et~al.}}},\ }\href@noop {} {\bibfield  {journal} {\bibinfo
  {journal} {Nucl.\ Phys.\ A}\ }\textbf {\bibinfo {volume} {119}},\ \bibinfo
  {pages} {305} (\bibinfo {year} {1968})}\BibitemShut {NoStop}%
\bibitem [{\citenamefont {{B\"urger}{\it~et~al.}}(2005)}]{cr56_2}%
  \BibitemOpen
  \bibfield  {author} {\bibinfo {author} {\bibfnamefont {A.}~\bibnamefont
  {{B\"urger}{\it~et~al.}}},\ }\href@noop {} {\bibfield  {journal} {\bibinfo
  {journal} {Phys.\ Lett.\ B}\ }\textbf {\bibinfo {volume} {622}},\ \bibinfo
  {pages} {29} (\bibinfo {year} {2005})}\BibitemShut {NoStop}%
\bibitem [{\citenamefont {{Steppenbeck}{\it~et~al.}}(2013)}]{gxpf1br}%
  \BibitemOpen
  \bibfield  {author} {\bibinfo {author} {\bibfnamefont {D.}~\bibnamefont
  {{Steppenbeck}{\it~et~al.}}},\ }\href@noop {} {\bibfield  {journal} {\bibinfo
   {journal} {Nature}\ }\textbf {\bibinfo {volume} {502}},\ \bibinfo {pages}
  {207} (\bibinfo {year} {2013})}\BibitemShut {NoStop}%
\bibitem [{\citenamefont {{Otsuka}{\it~et~al.}}(2001)}]{magic}%
  \BibitemOpen
  \bibfield  {author} {\bibinfo {author} {\bibfnamefont {T.}~\bibnamefont
  {{Otsuka}{\it~et~al.}}},\ }\href@noop {} {\bibfield  {journal} {\bibinfo
  {journal} {Phys.\ Rev.\ Lett.}\ }\textbf {\bibinfo {volume} {87}},\ \bibinfo
  {pages} {082502} (\bibinfo {year} {2001})}\BibitemShut {NoStop}%
\bibitem [{\citenamefont {{Honma}{\it~et~al.}}(2004)}]{gxpf1}%
  \BibitemOpen
  \bibfield  {author} {\bibinfo {author} {\bibfnamefont {M.}~\bibnamefont
  {{Honma}{\it~et~al.}}},\ }\href@noop {} {\bibfield  {journal} {\bibinfo
  {journal} {Phys.\ Rev.\ C}\ }\textbf {\bibinfo {volume} {69}},\ \bibinfo
  {pages} {034335} (\bibinfo {year} {2004})}\BibitemShut {NoStop}%
\bibitem [{\citenamefont {{Caurier}{\it~et~al.}}(2005)}]{Caurier_RMP}%
  \BibitemOpen
  \bibfield  {author} {\bibinfo {author} {\bibfnamefont {E.}~\bibnamefont
  {{Caurier}{\it~et~al.}}},\ }\href@noop {} {\bibfield  {journal} {\bibinfo
  {journal} {Rev.\ Mod.\ Phys}\ }\textbf {\bibinfo {volume} {77}},\ \bibinfo
  {pages} {427} (\bibinfo {year} {2005})}\BibitemShut {NoStop}%
\bibitem [{\citenamefont {{Otsuka}{\it~et~al.}}(2005)}]{tensor}%
  \BibitemOpen
  \bibfield  {author} {\bibinfo {author} {\bibfnamefont {T.}~\bibnamefont
  {{Otsuka}{\it~et~al.}}},\ }\href@noop {} {\bibfield  {journal} {\bibinfo
  {journal} {Phys.\ Rev.\ Lett.}\ }\textbf {\bibinfo {volume} {95}},\ \bibinfo
  {pages} {232502} (\bibinfo {year} {2005})}\BibitemShut {NoStop}%
\bibitem [{\citenamefont {{Broda}{\it~et~al.}}(1995)}]{ni68_broda}%
  \BibitemOpen
  \bibfield  {author} {\bibinfo {author} {\bibfnamefont {R.}~\bibnamefont
  {{Broda}{\it~et~al.}}},\ }\href@noop {} {\bibfield  {journal} {\bibinfo
  {journal} {Phys.\ Rev.\ Lett}\ }\textbf {\bibinfo {volume} {74}},\ \bibinfo
  {pages} {868} (\bibinfo {year} {1995})}\BibitemShut {NoStop}%
\bibitem [{\citenamefont {{Ishii}{\it~et~al.}}(2000)}]{ni68_ishii}%
  \BibitemOpen
  \bibfield  {author} {\bibinfo {author} {\bibfnamefont {T.}~\bibnamefont
  {{Ishii}{\it~et~al.}}},\ }\href@noop {} {\bibfield  {journal} {\bibinfo
  {journal} {Phys.\ Rev.\ Lett}\ }\textbf {\bibinfo {volume} {84}},\ \bibinfo
  {pages} {39} (\bibinfo {year} {2000})}\BibitemShut {NoStop}%
\bibitem [{\citenamefont {{Sorlin}{\it~et~al.}}(2002)}]{ni68_sorlin}%
  \BibitemOpen
  \bibfield  {author} {\bibinfo {author} {\bibfnamefont {O.}~\bibnamefont
  {{Sorlin}{\it~et~al.}}},\ }\href@noop {} {\bibfield  {journal} {\bibinfo
  {journal} {Phys.\ Rev.\ Lett}\ }\textbf {\bibinfo {volume} {88}},\ \bibinfo
  {pages} {092501} (\bibinfo {year} {2002})}\BibitemShut {NoStop}%
\bibitem [{\citenamefont {{Pritychenko}{\it~et~al.}}(2012)}]{cr_fe_2p}%
  \BibitemOpen
  \bibfield  {author} {\bibinfo {author} {\bibfnamefont {B.}~\bibnamefont
  {{Pritychenko}{\it~et~al.}}},\ }\href@noop {} {\bibfield  {journal} {\bibinfo
   {journal} {At.\ Data\ Nucl.\ Data\ Tables}\ }\textbf {\bibinfo {volume}
  {98}},\ \bibinfo {pages} {798} (\bibinfo {year} {2012})}\BibitemShut
  {NoStop}%
\bibitem [{\citenamefont {{Aoi}{\it~et~al.}}(2009)}]{cr60-62_1}%
  \BibitemOpen
  \bibfield  {author} {\bibinfo {author} {\bibfnamefont {N.}~\bibnamefont
  {{Aoi}{\it~et~al.}}},\ }\href@noop {} {\bibfield  {journal} {\bibinfo
  {journal} {Phys.\ Rev.\ Lett.}\ }\textbf {\bibinfo {volume} {102}},\ \bibinfo
  {pages} {012502} (\bibinfo {year} {2009})}\BibitemShut {NoStop}%
\bibitem [{\citenamefont {{Baugher}{\it~et~al.}}(2012)}]{cr60-62_2}%
  \BibitemOpen
  \bibfield  {author} {\bibinfo {author} {\bibfnamefont {T.}~\bibnamefont
  {{Baugher}{\it~et~al.}}},\ }\href@noop {} {\bibfield  {journal} {\bibinfo
  {journal} {Phys.\ Rev.\ C}\ }\textbf {\bibinfo {volume} {86}},\ \bibinfo
  {pages} {011305(R)} (\bibinfo {year} {2012})}\BibitemShut {NoStop}%
\bibitem [{\citenamefont {{Crawford}{\it~et~al.}}(2013)}]{cr64_fe68}%
  \BibitemOpen
  \bibfield  {author} {\bibinfo {author} {\bibfnamefont {H.~L.}\ \bibnamefont
  {{Crawford}{\it~et~al.}}},\ }\href@noop {} {\bibfield  {journal} {\bibinfo
  {journal} {Phys.\ Rev.\ Lett.}\ }\textbf {\bibinfo {volume} {110}},\ \bibinfo
  {pages} {242701} (\bibinfo {year} {2013})}\BibitemShut {NoStop}%
\bibitem [{\citenamefont {{Rother}{\it~et~al.}}(2011)}]{fe66}%
  \BibitemOpen
  \bibfield  {author} {\bibinfo {author} {\bibfnamefont {W.}~\bibnamefont
  {{Rother}{\it~et~al.}}},\ }\href@noop {} {\bibfield  {journal} {\bibinfo
  {journal} {Phys.\ Rev.\ Lett.}\ }\textbf {\bibinfo {volume} {106}},\ \bibinfo
  {pages} {022502} (\bibinfo {year} {2011})}\BibitemShut {NoStop}%
\bibitem [{\citenamefont {{Kaneko}{\it~et~al.}}(2008)}]{Kaneko}%
  \BibitemOpen
  \bibfield  {author} {\bibinfo {author} {\bibfnamefont {K.}~\bibnamefont
  {{Kaneko}{\it~et~al.}}},\ }\href@noop {} {\bibfield  {journal} {\bibinfo
  {journal} {Phys.\ Rev.\ C}\ }\textbf {\bibinfo {volume} {78}},\ \bibinfo
  {pages} {064312} (\bibinfo {year} {2008})}\BibitemShut {NoStop}%
\bibitem [{\citenamefont {Oba}\ and\ \citenamefont {Matsuo}(2008)}]{Oba}%
  \BibitemOpen
  \bibfield  {author} {\bibinfo {author} {\bibfnamefont {H.}~\bibnamefont
  {Oba}}\ and\ \bibinfo {author} {\bibfnamefont {M.}~\bibnamefont {Matsuo}},\
  }\href@noop {} {\bibfield  {journal} {\bibinfo  {journal} {Prog.\ Theor.\
  Phys.}\ }\textbf {\bibinfo {volume} {120}},\ \bibinfo {pages} {143} (\bibinfo
  {year} {2008})}\BibitemShut {NoStop}%
\bibitem [{\citenamefont {{Lenzi}{\it~et~al.}}(2010)}]{Lenzi}%
  \BibitemOpen
  \bibfield  {author} {\bibinfo {author} {\bibfnamefont {S.~M.}\ \bibnamefont
  {{Lenzi}{\it~et~al.}}},\ }\href@noop {} {\bibfield  {journal} {\bibinfo
  {journal} {Phys.\ Rev.\ C}\ }\textbf {\bibinfo {volume} {82}},\ \bibinfo
  {pages} {054301} (\bibinfo {year} {2010})}\BibitemShut {NoStop}%
\bibitem [{\citenamefont {{Gade}{\it~et~al.}}(2014)}]{ti}%
  \BibitemOpen
  \bibfield  {author} {\bibinfo {author} {\bibfnamefont {A.}~\bibnamefont
  {{Gade}{\it~et~al.}}},\ }\href@noop {} {\bibfield  {journal} {\bibinfo
  {journal} {Phys.\ Rev.\ Lett}\ }\textbf {\bibinfo {volume} {112}},\ \bibinfo
  {pages} {112503} (\bibinfo {year} {2014})}\BibitemShut {NoStop}%
\bibitem [{\citenamefont {{Warburton}{\it~et~al.}}(1990)}]{island}%
  \BibitemOpen
  \bibfield  {author} {\bibinfo {author} {\bibfnamefont {E.~K.}\ \bibnamefont
  {{Warburton}{\it~et~al.}}},\ }\href@noop {} {\bibfield  {journal} {\bibinfo
  {journal} {Phys.\ Rev.\ C}\ }\textbf {\bibinfo {volume} {41}},\ \bibinfo
  {pages} {1147} (\bibinfo {year} {1990})}\BibitemShut {NoStop}%
\bibitem [{\citenamefont {{Utsuno}{\it~et~al.}}(1999)}]{n20mcsm}%
  \BibitemOpen
  \bibfield  {author} {\bibinfo {author} {\bibfnamefont {Y.}~\bibnamefont
  {{Utsuno}{\it~et~al.}}},\ }\href@noop {} {\bibfield  {journal} {\bibinfo
  {journal} {Phys.\ Rev.\ C}\ }\textbf {\bibinfo {volume} {60}},\ \bibinfo
  {pages} {054315} (\bibinfo {year} {1999})}\BibitemShut {NoStop}%
\bibitem [{\citenamefont {{Tsunoda}{\it~et~al.}}(2014)}]{Tsunoda}%
  \BibitemOpen
  \bibfield  {author} {\bibinfo {author} {\bibfnamefont {Y.}~\bibnamefont
  {{Tsunoda}{\it~et~al.}}},\ }\href@noop {} {\bibfield  {journal} {\bibinfo
  {journal} {Phys.\ Rev.\ C}\ }\textbf {\bibinfo {volume} {89}},\ \bibinfo
  {pages} {031301(R)} (\bibinfo {year} {2014})}\BibitemShut {NoStop}%
\bibitem [{\citenamefont {{Appelbe}{\it~et~al.}}(2003)}]{cr55}%
  \BibitemOpen
  \bibfield  {author} {\bibinfo {author} {\bibfnamefont {D.~E.}\ \bibnamefont
  {{Appelbe}{\it~et~al.}}},\ }\href@noop {} {\bibfield  {journal} {\bibinfo
  {journal} {Phys.\ Rev.\ C}\ }\textbf {\bibinfo {volume} {67}},\ \bibinfo
  {pages} {034309} (\bibinfo {year} {2003})}\BibitemShut {NoStop}%
\bibitem [{\citenamefont {{Deacon}{\it~et~al.}}(2005)}]{cr57}%
  \BibitemOpen
  \bibfield  {author} {\bibinfo {author} {\bibfnamefont {A.~N.}\ \bibnamefont
  {{Deacon}{\it~et~al.}}},\ }\href@noop {} {\bibfield  {journal} {\bibinfo
  {journal} {Phys.\ Lett.\ B}\ }\textbf {\bibinfo {volume} {622}},\ \bibinfo
  {pages} {151} (\bibinfo {year} {2005})}\BibitemShut {NoStop}%
\bibitem [{\citenamefont {{Freeman}{\it~et~al.}}(2004)}]{cr59}%
  \BibitemOpen
  \bibfield  {author} {\bibinfo {author} {\bibfnamefont {S.~J.}\ \bibnamefont
  {{Freeman}{\it~et~al.}}},\ }\href@noop {} {\bibfield  {journal} {\bibinfo
  {journal} {Phys.\ Rev.\ C}\ }\textbf {\bibinfo {volume} {69}},\ \bibinfo
  {pages} {064301} (\bibinfo {year} {2004})}\BibitemShut {NoStop}%
\bibitem [{\citenamefont {{Deacon}{\it~et~al.}}(2007)}]{fe59-60}%
  \BibitemOpen
  \bibfield  {author} {\bibinfo {author} {\bibfnamefont {A.~N.}\ \bibnamefont
  {{Deacon}{\it~et~al.}}},\ }\href@noop {} {\bibfield  {journal} {\bibinfo
  {journal} {Phys.\ Rev.\ C}\ }\textbf {\bibinfo {volume} {76}},\ \bibinfo
  {pages} {054303} (\bibinfo {year} {2007})}\BibitemShut {NoStop}%
\bibitem [{\citenamefont {{Hoteling}{\it~et~al.}}(2008)}]{fe61}%
  \BibitemOpen
  \bibfield  {author} {\bibinfo {author} {\bibfnamefont {N.}~\bibnamefont
  {{Hoteling}{\it~et~al.}}},\ }\href@noop {} {\bibfield  {journal} {\bibinfo
  {journal} {Phys.\ Rev.\ C}\ }\textbf {\bibinfo {volume} {77}},\ \bibinfo
  {pages} {044314} (\bibinfo {year} {2008})}\BibitemShut {NoStop}%
\bibitem [{\citenamefont {{Deacon}{\it~et~al.}}(2011)}]{cr55_2}%
  \BibitemOpen
  \bibfield  {author} {\bibinfo {author} {\bibfnamefont {A.~N.}\ \bibnamefont
  {{Deacon}{\it~et~al.}}},\ }\href@noop {} {\bibfield  {journal} {\bibinfo
  {journal} {Phys.\ Rev.\ C}\ }\textbf {\bibinfo {volume} {83}},\ \bibinfo
  {pages} {064305} (\bibinfo {year} {2011})}\BibitemShut {NoStop}%
\bibitem [{\citenamefont {{Lalazissis}{\it~et~al.}}(1998)}]{Lalazissis}%
  \BibitemOpen
  \bibfield  {author} {\bibinfo {author} {\bibfnamefont {G.~A.}\ \bibnamefont
  {{Lalazissis}{\it~et~al.}}},\ }\href@noop {} {\bibfield  {journal} {\bibinfo
  {journal} {Nucl.\ Phys.\ A}\ }\textbf {\bibinfo {volume} {628}},\ \bibinfo
  {pages} {221} (\bibinfo {year} {1998})}\BibitemShut {NoStop}%
\bibitem [{\citenamefont {{Yang}{\it~et~al.}}(2011)}]{cr59_def}%
  \BibitemOpen
  \bibfield  {author} {\bibinfo {author} {\bibfnamefont {Y.-C.}\ \bibnamefont
  {{Yang}{\it~et~al.}}},\ }\href@noop {} {\bibfield  {journal} {\bibinfo
  {journal} {Phys.\ Lett.\ B}\ }\textbf {\bibinfo {volume} {700}},\ \bibinfo
  {pages} {44} (\bibinfo {year} {2011})}\BibitemShut {NoStop}%
\bibitem [{\citenamefont {{Matea}{\it~et~al.}}(2004)}]{fe61_g}%
  \BibitemOpen
  \bibfield  {author} {\bibinfo {author} {\bibfnamefont {I.}~\bibnamefont
  {{Matea}{\it~et~al.}}},\ }\href@noop {} {\bibfield  {journal} {\bibinfo
  {journal} {Phys.\ Rev.\ Lett.}\ }\textbf {\bibinfo {volume} {93}},\ \bibinfo
  {pages} {142503} (\bibinfo {year} {2004})}\BibitemShut {NoStop}%
\bibitem [{\citenamefont {{Vermeulen}{\it~et~al.}}(2007)}]{fe61_sqm}%
  \BibitemOpen
  \bibfield  {author} {\bibinfo {author} {\bibfnamefont {N.}~\bibnamefont
  {{Vermeulen}{\it~et~al.}}},\ }\href@noop {} {\bibfield  {journal} {\bibinfo
  {journal} {Phys.\ Rev.\ C}\ }\textbf {\bibinfo {volume} {75}},\ \bibinfo
  {pages} {051302(R)} (\bibinfo {year} {2007})}\BibitemShut {NoStop}%
\bibitem [{\citenamefont {{Zuker}{\it~et~al.}}(1995)}]{q_SU3}%
  \BibitemOpen
  \bibfield  {author} {\bibinfo {author} {\bibfnamefont {A.~P.}\ \bibnamefont
  {{Zuker}{\it~et~al.}}},\ }\href@noop {} {\bibfield  {journal} {\bibinfo
  {journal} {Phys.\ Rev.\ C}\ }\textbf {\bibinfo {volume} {52}},\ \bibinfo
  {pages} {R1741} (\bibinfo {year} {1995})}\BibitemShut {NoStop}%
\bibitem [{\citenamefont {Gloeckner}\ and\ \citenamefont
  {Lawson}(1974)}]{Lawson}%
  \BibitemOpen
  \bibfield  {author} {\bibinfo {author} {\bibfnamefont {D.~H.}\ \bibnamefont
  {Gloeckner}}\ and\ \bibinfo {author} {\bibfnamefont {R.~D.}\ \bibnamefont
  {Lawson}},\ }\href@noop {} {\bibfield  {journal} {\bibinfo  {journal} {Phys.\
  Lett.}\ }\textbf {\bibinfo {volume} {53B}},\ \bibinfo {pages} {313} (\bibinfo
  {year} {1974})}\BibitemShut {NoStop}%
\bibitem [{\citenamefont {Mizusaki}\ \emph {et~al.}()\citenamefont {Mizusaki},
  \citenamefont {Shimizu}, \citenamefont {Utsuno},\ and\ \citenamefont
  {Honma}}]{MSHELL64}%
  \BibitemOpen
  \bibfield  {author} {\bibinfo {author} {\bibfnamefont {T.}~\bibnamefont
  {Mizusaki}}, \bibinfo {author} {\bibfnamefont {N.}~\bibnamefont {Shimizu}},
  \bibinfo {author} {\bibfnamefont {Y.}~\bibnamefont {Utsuno}}, \ and\ \bibinfo
  {author} {\bibfnamefont {M.}~\bibnamefont {Honma}},\ }\href@noop {}
  {}\bibinfo {howpublished} {private communications}\BibitemShut {NoStop}%
\bibitem [{\citenamefont {Shimizu}()}]{KSHELL}%
  \BibitemOpen
  \bibfield  {author} {\bibinfo {author} {\bibfnamefont {N.}~\bibnamefont
  {Shimizu}},\ }\href@noop {} {}\bibinfo {howpublished}
  {arXiv:1310.5431}\BibitemShut {NoStop}%
\bibitem [{\citenamefont {{Honma}{\it~et~al.}}(2005)}]{gxpf1a}%
  \BibitemOpen
  \bibfield  {author} {\bibinfo {author} {\bibfnamefont {M.}~\bibnamefont
  {{Honma}{\it~et~al.}}},\ }\href@noop {} {\bibfield  {journal} {\bibinfo
  {journal} {Eur.\ Phys.\ J.\ A}\ }\textbf {\bibinfo {volume} {25}},\ \bibinfo
  {pages} {s01, 499} (\bibinfo {year} {2005})}\BibitemShut {NoStop}%
\bibitem [{\citenamefont {{Honma}{\it~et~al.}}(2008)}]{gxpf1b}%
  \BibitemOpen
  \bibfield  {author} {\bibinfo {author} {\bibfnamefont {M.}~\bibnamefont
  {{Honma}{\it~et~al.}}},\ }\href@noop {} {\bibfield  {journal} {\bibinfo
  {journal} {RIKEN Accelerator Progress Report}\ }\textbf {\bibinfo {volume}
  {41}},\ \bibinfo {pages} {32} (\bibinfo {year} {2008})}\BibitemShut {NoStop}%
\bibitem [{\citenamefont {{Otsuka}{\it~et~al.}}(2010)}]{vmu}%
  \BibitemOpen
  \bibfield  {author} {\bibinfo {author} {\bibfnamefont {T.}~\bibnamefont
  {{Otsuka}{\it~et~al.}}},\ }\href@noop {} {\bibfield  {journal} {\bibinfo
  {journal} {Phys.\ Rev.\ Lett.}\ }\textbf {\bibinfo {volume} {104}},\ \bibinfo
  {pages} {012501} (\bibinfo {year} {2010})}\BibitemShut {NoStop}%
\bibitem [{\citenamefont {{Utsuno}{\it~et~al.}}(2012)}]{vmu_r}%
  \BibitemOpen
  \bibfield  {author} {\bibinfo {author} {\bibfnamefont {Y.}~\bibnamefont
  {{Utsuno}{\it~et~al.}}},\ }\href@noop {} {\bibfield  {journal} {\bibinfo
  {journal} {Phys.\ Rev.\ C}\ }\textbf {\bibinfo {volume} {86}},\ \bibinfo
  {pages} {051301(R)} (\bibinfo {year} {2012})}\BibitemShut {NoStop}%
\bibitem [{\citenamefont {{Bertsch}{\it~et~al.}}(1977)}]{M3Y}%
  \BibitemOpen
  \bibfield  {author} {\bibinfo {author} {\bibfnamefont {G.}~\bibnamefont
  {{Bertsch}{\it~et~al.}}},\ }\href@noop {} {\bibfield  {journal} {\bibinfo
  {journal} {Nucl.\ Phys.\ A}\ }\textbf {\bibinfo {volume} {284}},\ \bibinfo
  {pages} {399} (\bibinfo {year} {1977})}\BibitemShut {NoStop}%
\bibitem [{\citenamefont {{Brown}{\it~et~al.}}(1988)}]{Brown}%
  \BibitemOpen
  \bibfield  {author} {\bibinfo {author} {\bibfnamefont {B.~A.}\ \bibnamefont
  {{Brown}{\it~et~al.}}},\ }\href@noop {} {\bibfield  {journal} {\bibinfo
  {journal} {Ann.\ Phys.}\ }\textbf {\bibinfo {volume} {182}},\ \bibinfo
  {pages} {191} (\bibinfo {year} {1988})}\BibitemShut {NoStop}%
\bibitem [{\citenamefont {{Junde}}(2008)}]{NDS_A55}%
  \BibitemOpen
  \bibfield  {author} {\bibinfo {author} {\bibfnamefont {H.}~\bibnamefont
  {{Junde}}},\ }\href@noop {} {\bibfield  {journal} {\bibinfo  {journal}
  {Nucl.\ Data\ Sheets}\ }\textbf {\bibinfo {volume} {109}},\ \bibinfo {pages}
  {787} (\bibinfo {year} {2008})},\ \bibinfo {note} {and references
  therein}\BibitemShut {NoStop}%
\bibitem [{\citenamefont {{Bhat}}(1998)}]{NDS_A57}%
  \BibitemOpen
  \bibfield  {author} {\bibinfo {author} {\bibfnamefont {M.~R.}\ \bibnamefont
  {{Bhat}}},\ }\href@noop {} {\bibfield  {journal} {\bibinfo  {journal} {Nucl.\
  Data\ Sheets}\ }\textbf {\bibinfo {volume} {85}},\ \bibinfo {pages} {415}
  (\bibinfo {year} {1998})},\ \bibinfo {note} {and references
  therein}\BibitemShut {NoStop}%
\bibitem [{\citenamefont {{Baglin}}(2002)}]{NDS_A59}%
  \BibitemOpen
  \bibfield  {author} {\bibinfo {author} {\bibfnamefont {C.~M.}\ \bibnamefont
  {{Baglin}}},\ }\href@noop {} {\bibfield  {journal} {\bibinfo  {journal}
  {Nucl.\ Data\ Sheets}\ }\textbf {\bibinfo {volume} {95}},\ \bibinfo {pages}
  {215} (\bibinfo {year} {2002})},\ \bibinfo {note} {and references
  therein}\BibitemShut {NoStop}%
\bibitem [{\citenamefont {Bohr}\ and\ \citenamefont {Mottelson}(1975)}]{Bohr}%
  \BibitemOpen
  \bibfield  {author} {\bibinfo {author} {\bibfnamefont {A.}~\bibnamefont
  {Bohr}}\ and\ \bibinfo {author} {\bibfnamefont {B.~R.}\ \bibnamefont
  {Mottelson}},\ }\href@noop {} {\emph {\bibinfo {title} {Nuclear Structure
  vol. 2}}}\ (\bibinfo {year} {1975})\BibitemShut {NoStop}%
\bibitem [{\citenamefont {{Zhu}{\it~et~al.}}(2006)}]{cr56_cr58}%
  \BibitemOpen
  \bibfield  {author} {\bibinfo {author} {\bibfnamefont {S.}~\bibnamefont
  {{Zhu}{\it~et~al.}}},\ }\href@noop {} {\bibfield  {journal} {\bibinfo
  {journal} {Phys.\ Rev.\ C}\ }\textbf {\bibinfo {volume} {74}},\ \bibinfo
  {pages} {064315} (\bibinfo {year} {2006})}\BibitemShut {NoStop}%
\bibitem [{\citenamefont {{Steppenbeck}{\it~et~al.}}(2012)}]{fe58}%
  \BibitemOpen
  \bibfield  {author} {\bibinfo {author} {\bibfnamefont {D.}~\bibnamefont
  {{Steppenbeck}{\it~et~al.}}},\ }\href@noop {} {\bibfield  {journal} {\bibinfo
   {journal} {Phys.\ Rev.\ C}\ }\textbf {\bibinfo {volume} {85}},\ \bibinfo
  {pages} {044316} (\bibinfo {year} {2012})}\BibitemShut {NoStop}%
\bibitem [{\citenamefont {{Junde}{\it~et~al.}}(2011)}]{NDS_A56}%
  \BibitemOpen
  \bibfield  {author} {\bibinfo {author} {\bibfnamefont {H.}~\bibnamefont
  {{Junde}{\it~et~al.}}},\ }\href@noop {} {\bibfield  {journal} {\bibinfo
  {journal} {Nucl.\ Data\ Sheets}\ }\textbf {\bibinfo {volume} {112}},\
  \bibinfo {pages} {1513} (\bibinfo {year} {2011})},\ \bibinfo {note} {and
  references therein}\BibitemShut {NoStop}%
\bibitem [{\citenamefont {{Nesaraja}{\it~et~al.}}(2010)}]{NDS_A58}%
  \BibitemOpen
  \bibfield  {author} {\bibinfo {author} {\bibfnamefont {C.~D.}\ \bibnamefont
  {{Nesaraja}{\it~et~al.}}},\ }\href@noop {} {\bibfield  {journal} {\bibinfo
  {journal} {Nucl.\ Data\ Sheets}\ }\textbf {\bibinfo {volume} {111}},\
  \bibinfo {pages} {897} (\bibinfo {year} {2010})},\ \bibinfo {note} {and
  references therein}\BibitemShut {NoStop}%
\bibitem [{\citenamefont {Browne}\ and\ \citenamefont {Tuli}(2013)}]{NDS_A60}%
  \BibitemOpen
  \bibfield  {author} {\bibinfo {author} {\bibfnamefont {E.}~\bibnamefont
  {Browne}}\ and\ \bibinfo {author} {\bibfnamefont {J.~K.}\ \bibnamefont
  {Tuli}},\ }\href@noop {} {\bibfield  {journal} {\bibinfo  {journal} {Nucl.\
  Data\ Sheets}\ }\textbf {\bibinfo {volume} {114}},\ \bibinfo {pages} {1849}
  (\bibinfo {year} {2013})},\ \bibinfo {note} {and references
  therein}\BibitemShut {NoStop}%
\bibitem [{\citenamefont {{Sun}{\it~et~al.}}(2009)}]{PSM_fe58}%
  \BibitemOpen
  \bibfield  {author} {\bibinfo {author} {\bibfnamefont {Y.}~\bibnamefont
  {{Sun}{\it~et~al.}}},\ }\href@noop {} {\bibfield  {journal} {\bibinfo
  {journal} {Phys.\ Rev.\ C}\ }\textbf {\bibinfo {volume} {80}},\ \bibinfo
  {pages} {054306} (\bibinfo {year} {2009})}\BibitemShut {NoStop}%
\bibitem [{\citenamefont {Ring}\ and\ \citenamefont {Schuck}(1980)}]{Ring}%
  \BibitemOpen
  \bibfield  {author} {\bibinfo {author} {\bibfnamefont {P.}~\bibnamefont
  {Ring}}\ and\ \bibinfo {author} {\bibfnamefont {P.}~\bibnamefont {Schuck}},\
  }\href@noop {} {\emph {\bibinfo {title} {The Nuclear Many-Body Problem}}}\
  (\bibinfo {year} {1980})\BibitemShut {NoStop}%
\bibitem [{\citenamefont {Stephens}(1975)}]{part_rot}%
  \BibitemOpen
  \bibfield  {author} {\bibinfo {author} {\bibfnamefont {F.~S.}\ \bibnamefont
  {Stephens}},\ }\href@noop {} {\bibfield  {journal} {\bibinfo  {journal}
  {Rev.\ Mod.\ Phys.}\ }\textbf {\bibinfo {volume} {47}},\ \bibinfo {pages}
  {43} (\bibinfo {year} {1975})}\BibitemShut {NoStop}%
\bibitem [{\citenamefont {{Mizusaki}{\it~et~al.}}(1999)}]{ni56}%
  \BibitemOpen
  \bibfield  {author} {\bibinfo {author} {\bibfnamefont {T.}~\bibnamefont
  {{Mizusaki}{\it~et~al.}}},\ }\href@noop {} {\bibfield  {journal} {\bibinfo
  {journal} {Phys.\ Rev.\ C}\ }\textbf {\bibinfo {volume} {59}},\ \bibinfo
  {pages} {R1846} (\bibinfo {year} {1999})}\BibitemShut {NoStop}%
\bibitem [{\citenamefont {{Shimizu}{\it~et~al.}}(2012)}]{MCSM}%
  \BibitemOpen
  \bibfield  {author} {\bibinfo {author} {\bibfnamefont {N.}~\bibnamefont
  {{Shimizu}{\it~et~al.}}},\ }\href@noop {} {\bibfield  {journal} {\bibinfo
  {journal} {Phys.\ Rev.\ C}\ }\textbf {\bibinfo {volume} {85}},\ \bibinfo
  {pages} {054301} (\bibinfo {year} {2012})}\BibitemShut {NoStop}%
\bibitem [{\citenamefont {{Cavallaro}{\it~et~al.}}(1977)}]{fe58_old}%
  \BibitemOpen
  \bibfield  {author} {\bibinfo {author} {\bibfnamefont {S.}~\bibnamefont
  {{Cavallaro}{\it~et~al.}}},\ }\href@noop {} {\bibfield  {journal} {\bibinfo
  {journal} {Nucl.\ Phys.\ A}\ }\textbf {\bibinfo {volume} {293}},\ \bibinfo
  {pages} {125} (\bibinfo {year} {1977})}\BibitemShut {NoStop}%
\bibitem [{\citenamefont {{Junde}}(2009)}]{NDS_A53}%
  \BibitemOpen
  \bibfield  {author} {\bibinfo {author} {\bibfnamefont {H.}~\bibnamefont
  {{Junde}}},\ }\href@noop {} {\bibfield  {journal} {\bibinfo  {journal}
  {Nucl.\ Data\ Sheets}\ }\textbf {\bibinfo {volume} {110}},\ \bibinfo {pages}
  {2689} (\bibinfo {year} {2009})},\ \bibinfo {note} {and references
  therein}\BibitemShut {NoStop}%
\bibitem [{\citenamefont {{Bhat}}(1999)}]{NDS_A61}%
  \BibitemOpen
  \bibfield  {author} {\bibinfo {author} {\bibfnamefont {M.~R.}\ \bibnamefont
  {{Bhat}}},\ }\href@noop {} {\bibfield  {journal} {\bibinfo  {journal} {Nucl.\
  Data\ Sheets}\ }\textbf {\bibinfo {volume} {88}},\ \bibinfo {pages} {417}
  (\bibinfo {year} {1999})},\ \bibinfo {note} {and references
  therein}\BibitemShut {NoStop}%
\bibitem [{\citenamefont {{B.~Erjun and H.~Junde}}(2001)}]{NDS_A63}%
  \BibitemOpen
  \bibfield  {author} {\bibinfo {author} {\bibnamefont {{B.~Erjun and
  H.~Junde}}},\ }\href@noop {} {\bibfield  {journal} {\bibinfo  {journal}
  {Nucl.\ Data\ Sheets}\ }\textbf {\bibinfo {volume} {92}},\ \bibinfo {pages}
  {147} (\bibinfo {year} {2001})},\ \bibinfo {note} {and references
  therein}\BibitemShut {NoStop}%
\bibitem [{\citenamefont {{Rao}{\it~et~al.}}(1968)}]{53Cr_dp}%
  \BibitemOpen
  \bibfield  {author} {\bibinfo {author} {\bibfnamefont {M.~N.}\ \bibnamefont
  {{Rao}{\it~et~al.}}},\ }\href@noop {} {\bibfield  {journal} {\bibinfo
  {journal} {Nucl.\ Phys.\ A}\ }\textbf {\bibinfo {volume} {121}},\ \bibinfo
  {pages} {1} (\bibinfo {year} {1968})}\BibitemShut {NoStop}%
\bibitem [{\citenamefont {{Bock}{\it~et~al.}}(1965)}]{53Cr55Cr_dp}%
  \BibitemOpen
  \bibfield  {author} {\bibinfo {author} {\bibfnamefont {R.}~\bibnamefont
  {{Bock}{\it~et~al.}}},\ }\href@noop {} {\bibfield  {journal} {\bibinfo
  {journal} {Nucl.\ Phys.}\ }\textbf {\bibinfo {volume} {72}},\ \bibinfo
  {pages} {273} (\bibinfo {year} {1965})}\BibitemShut {NoStop}%
\bibitem [{\citenamefont {Kocher}\ and\ \citenamefont
  {Haeberli}(1972)}]{53Cr55Fe_dp}%
  \BibitemOpen
  \bibfield  {author} {\bibinfo {author} {\bibfnamefont {D.~C.}\ \bibnamefont
  {Kocher}}\ and\ \bibinfo {author} {\bibfnamefont {W.}~\bibnamefont
  {Haeberli}},\ }\href@noop {} {\bibfield  {journal} {\bibinfo  {journal}
  {Nucl.\ Phys.\ A}\ }\textbf {\bibinfo {volume} {196}},\ \bibinfo {pages}
  {225} (\bibinfo {year} {1972})}\BibitemShut {NoStop}%
\bibitem [{\citenamefont {Macgregor}\ and\ \citenamefont
  {Brown}(1972)}]{55Cr_dp}%
  \BibitemOpen
  \bibfield  {author} {\bibinfo {author} {\bibfnamefont {A.~E.}\ \bibnamefont
  {Macgregor}}\ and\ \bibinfo {author} {\bibfnamefont {G.}~\bibnamefont
  {Brown}},\ }\href@noop {} {\bibfield  {journal} {\bibinfo  {journal} {Nucl.\
  Phys.\ A}\ }\textbf {\bibinfo {volume} {198}},\ \bibinfo {pages} {237}
  (\bibinfo {year} {1972})}\BibitemShut {NoStop}%
\bibitem [{\citenamefont {Fulmer}\ and\ \citenamefont
  {McCarthy}(1963)}]{55Fe63Ni_dp}%
  \BibitemOpen
  \bibfield  {author} {\bibinfo {author} {\bibfnamefont {R.~H.}\ \bibnamefont
  {Fulmer}}\ and\ \bibinfo {author} {\bibfnamefont {A.~L.}\ \bibnamefont
  {McCarthy}},\ }\href@noop {} {\bibfield  {journal} {\bibinfo  {journal}
  {Phys.\ Rev.}\ }\textbf {\bibinfo {volume} {131}},\ \bibinfo {pages} {2133}
  (\bibinfo {year} {1963})}\BibitemShut {NoStop}%
\bibitem [{\citenamefont {Maxwell}\ and\ \citenamefont
  {Parkinson}(1964)}]{55Fe_dp}%
  \BibitemOpen
  \bibfield  {author} {\bibinfo {author} {\bibfnamefont {J.~R.}\ \bibnamefont
  {Maxwell}}\ and\ \bibinfo {author} {\bibfnamefont {W.~C.}\ \bibnamefont
  {Parkinson}},\ }\href@noop {} {\bibfield  {journal} {\bibinfo  {journal}
  {Phys.\ Rev.}\ }\textbf {\bibinfo {volume} {135}},\ \bibinfo {pages} {B82}
  (\bibinfo {year} {1964})}\BibitemShut {NoStop}%
\bibitem [{\citenamefont {{Sen~Gupta}{\it~et~al.}}(1971)}]{57Fe_dp1}%
  \BibitemOpen
  \bibfield  {author} {\bibinfo {author} {\bibfnamefont {H.~M.}\ \bibnamefont
  {{Sen~Gupta}{\it~et~al.}}},\ }\href@noop {} {\bibfield  {journal} {\bibinfo
  {journal} {Nucl.\ Phys.\ A}\ }\textbf {\bibinfo {volume} {160}},\ \bibinfo
  {pages} {529} (\bibinfo {year} {1971})}\BibitemShut {NoStop}%
\bibitem [{\citenamefont {Thomson}(1974)}]{57Fe_dp2}%
  \BibitemOpen
  \bibfield  {author} {\bibinfo {author} {\bibfnamefont {J.~A.}\ \bibnamefont
  {Thomson}},\ }\href@noop {} {\bibfield  {journal} {\bibinfo  {journal}
  {Nucl.\ Phys.\ A}\ }\textbf {\bibinfo {volume} {227}},\ \bibinfo {pages}
  {485} (\bibinfo {year} {1974})}\BibitemShut {NoStop}%
\bibitem [{\citenamefont {{McLean}{\it~et~al.}}(1972)}]{59Fe_dp}%
  \BibitemOpen
  \bibfield  {author} {\bibinfo {author} {\bibfnamefont {K.~C.}\ \bibnamefont
  {{McLean}{\it~et~al.}}},\ }\href@noop {} {\bibfield  {journal} {\bibinfo
  {journal} {Nucl.\ Phys.\ A}\ }\textbf {\bibinfo {volume} {191}},\ \bibinfo
  {pages} {417} (\bibinfo {year} {1972})}\BibitemShut {NoStop}%
\bibitem [{\citenamefont {Taylor}\ and\ \citenamefont
  {Cameron}(1980)}]{55Cr59Fe59Ni_dp}%
  \BibitemOpen
  \bibfield  {author} {\bibinfo {author} {\bibfnamefont {T.}~\bibnamefont
  {Taylor}}\ and\ \bibinfo {author} {\bibfnamefont {J.~A.}\ \bibnamefont
  {Cameron}},\ }\href@noop {} {\bibfield  {journal} {\bibinfo  {journal}
  {Nucl.\ Phys.\ A}\ }\textbf {\bibinfo {volume} {337}},\ \bibinfo {pages}
  {389} (\bibinfo {year} {1980})}\BibitemShut {NoStop}%
\bibitem [{\citenamefont {Chowdhury}\ and\ \citenamefont
  {{Sen~Gupta}}(1973)}]{59Ni_dp1}%
  \BibitemOpen
  \bibfield  {author} {\bibinfo {author} {\bibfnamefont {M.~S.}\ \bibnamefont
  {Chowdhury}}\ and\ \bibinfo {author} {\bibfnamefont {H.~M.}\ \bibnamefont
  {{Sen~Gupta}}},\ }\href@noop {} {\bibfield  {journal} {\bibinfo  {journal}
  {Nucl.\ Phys.\ A}\ }\textbf {\bibinfo {volume} {205}},\ \bibinfo {pages}
  {454} (\bibinfo {year} {1973})}\BibitemShut {NoStop}%
\bibitem [{\citenamefont {{Aymar}{\it~et~al.}}(1973{\natexlab{a}})}]{59Ni_dp2}%
  \BibitemOpen
  \bibfield  {author} {\bibinfo {author} {\bibfnamefont {J.~A.}\ \bibnamefont
  {{Aymar}{\it~et~al.}}},\ }\href@noop {} {\bibfield  {journal} {\bibinfo
  {journal} {Nucl.\ Phys.\ A}\ }\textbf {\bibinfo {volume} {207}},\ \bibinfo
  {pages} {596} (\bibinfo {year} {1973}{\natexlab{a}})}\BibitemShut {NoStop}%
\bibitem [{\citenamefont {{Iwamoto}{\it~et~al.}}(1994)}]{59Ni_dp3}%
  \BibitemOpen
  \bibfield  {author} {\bibinfo {author} {\bibfnamefont {O.}~\bibnamefont
  {{Iwamoto}{\it~et~al.}}},\ }\href@noop {} {\bibfield  {journal} {\bibinfo
  {journal} {Nucl.\ Phys.\ A}\ }\textbf {\bibinfo {volume} {576}},\ \bibinfo
  {pages} {387} (\bibinfo {year} {1994})}\BibitemShut {NoStop}%
\bibitem [{\citenamefont {{Cohen}{\it~et~al.}}(1962)}]{61Ni_dp1}%
  \BibitemOpen
  \bibfield  {author} {\bibinfo {author} {\bibfnamefont {B.~L.}\ \bibnamefont
  {{Cohen}{\it~et~al.}}},\ }\href@noop {} {\bibfield  {journal} {\bibinfo
  {journal} {Phys.\ Rev.}\ }\textbf {\bibinfo {volume} {126}},\ \bibinfo
  {pages} {698} (\bibinfo {year} {1962})}\BibitemShut {NoStop}%
\bibitem [{\citenamefont {{Fulmer}{\it~et~al.}}(1964)}]{61Ni_dp2}%
  \BibitemOpen
  \bibfield  {author} {\bibinfo {author} {\bibfnamefont {R.~H.}\ \bibnamefont
  {{Fulmer}{\it~et~al.}}},\ }\href@noop {} {\bibfield  {journal} {\bibinfo
  {journal} {Phys.\ Rev.}\ }\textbf {\bibinfo {volume} {133}},\ \bibinfo
  {pages} {B955} (\bibinfo {year} {1964})}\BibitemShut {NoStop}%
\bibitem [{\citenamefont {{Cosman}{\it~et~al.}}(1967)}]{61Ni_dp3}%
  \BibitemOpen
  \bibfield  {author} {\bibinfo {author} {\bibfnamefont {E.~R.}\ \bibnamefont
  {{Cosman}{\it~et~al.}}},\ }\href@noop {} {\bibfield  {journal} {\bibinfo
  {journal} {Phys.\ Rev.}\ }\textbf {\bibinfo {volume} {163}},\ \bibinfo
  {pages} {1134} (\bibinfo {year} {1967})}\BibitemShut {NoStop}%
\bibitem [{\citenamefont {{Aymar}{\it~et~al.}}(1973{\natexlab{b}})}]{61Ni_dp4}%
  \BibitemOpen
  \bibfield  {author} {\bibinfo {author} {\bibfnamefont {J.~A.}\ \bibnamefont
  {{Aymar}{\it~et~al.}}},\ }\href@noop {} {\bibfield  {journal} {\bibinfo
  {journal} {Nucl.\ Phys. A}\ }\textbf {\bibinfo {volume} {213}},\ \bibinfo
  {pages} {125} (\bibinfo {year} {1973}{\natexlab{b}})}\BibitemShut {NoStop}%
\bibitem [{\citenamefont {{Turkiewicz}{\it~et~al.}}(1970)}]{63Ni_dp1}%
  \BibitemOpen
  \bibfield  {author} {\bibinfo {author} {\bibfnamefont {I.~M.}\ \bibnamefont
  {{Turkiewicz}{\it~et~al.}}},\ }\href@noop {} {\bibfield  {journal} {\bibinfo
  {journal} {Nucl.\ Phys.\ A}\ }\textbf {\bibinfo {volume} {143}},\ \bibinfo
  {pages} {641} (\bibinfo {year} {1970})}\BibitemShut {NoStop}%
\bibitem [{\citenamefont {{Anfinsen}{\it~et~al.}}(1970)}]{63Ni_dp2}%
  \BibitemOpen
  \bibfield  {author} {\bibinfo {author} {\bibfnamefont {T.~R.}\ \bibnamefont
  {{Anfinsen}{\it~et~al.}}},\ }\href@noop {} {\bibfield  {journal} {\bibinfo
  {journal} {Nucl.\ Phys.\ A}\ }\textbf {\bibinfo {volume} {157}},\ \bibinfo
  {pages} {561} (\bibinfo {year} {1970})}\BibitemShut {NoStop}%
\bibitem [{\citenamefont {{Huttlin}{\it~et~al.}}(1974)}]{63Ni_dp3}%
  \BibitemOpen
  \bibfield  {author} {\bibinfo {author} {\bibfnamefont {G.~A.}\ \bibnamefont
  {{Huttlin}{\it~et~al.}}},\ }\href@noop {} {\bibfield  {journal} {\bibinfo
  {journal} {Nucl.\ Phys.\ A}\ }\textbf {\bibinfo {volume} {227}},\ \bibinfo
  {pages} {389} (\bibinfo {year} {1974})}\BibitemShut {NoStop}%
\bibitem [{\citenamefont {{Roussel}{\it~et~al.}}(1970)}]{55Fe59Ni_He43}%
  \BibitemOpen
  \bibfield  {author} {\bibinfo {author} {\bibfnamefont {P.}~\bibnamefont
  {{Roussel}{\it~et~al.}}},\ }\href@noop {} {\bibfield  {journal} {\bibinfo
  {journal} {Nucl.\ Phys.\ A}\ }\textbf {\bibinfo {volume} {155}},\ \bibinfo
  {pages} {306} (\bibinfo {year} {1970})}\BibitemShut {NoStop}%
\bibitem [{\citenamefont {{Karban}{\it~et~al.}}(1991)}]{55Fe_Li76}%
  \BibitemOpen
  \bibfield  {author} {\bibinfo {author} {\bibfnamefont {O.}~\bibnamefont
  {{Karban}{\it~et~al.}}},\ }\href@noop {} {\bibfield  {journal} {\bibinfo
  {journal} {Nucl.\ Phys.\ A}\ }\textbf {\bibinfo {volume} {535}},\ \bibinfo
  {pages} {377} (\bibinfo {year} {1991})}\BibitemShut {NoStop}%
\bibitem [{\citenamefont {{Videbaek}{\it~et~al.}}(1985)}]{59Ni_C1413}%
  \BibitemOpen
  \bibfield  {author} {\bibinfo {author} {\bibfnamefont {F.}~\bibnamefont
  {{Videbaek}{\it~et~al.}}},\ }\href@noop {} {\bibfield  {journal} {\bibinfo
  {journal} {Nucl.\ Phys.\ A}\ }\textbf {\bibinfo {volume} {433}},\ \bibinfo
  {pages} {441} (\bibinfo {year} {1985})}\BibitemShut {NoStop}%
\bibitem [{\citenamefont {{Storm}{\it~et~al.}}(1983)}]{ESPE}%
  \BibitemOpen
  \bibfield  {author} {\bibinfo {author} {\bibfnamefont {M.~H.}\ \bibnamefont
  {{Storm}{\it~et~al.}}},\ }\href@noop {} {\bibfield  {journal} {\bibinfo
  {journal} {J.\ Phys.\ G:\ Nucl.\ Phys.}\ }\textbf {\bibinfo {volume} {9}},\
  \bibinfo {pages} {L165} (\bibinfo {year} {1983})}\BibitemShut {NoStop}%
\end{thebibliography}%

\end{document}